\documentclass[aps,amsmath,amssymb,twocolumn,pra,nofootinbib]{revtex4-1} 

\usepackage{graphicx} 
\usepackage{color} 
\usepackage{soul} 
\usepackage{lmodern} 
\usepackage{lmodern} 
\usepackage{amsmath} 
\usepackage{dsfont} 
\usepackage{placeins} 
\usepackage{amsthm} 
\usepackage{amsfonts} 
\usepackage{natbib}
\usepackage[english]{babel}
\usepackage{comment}

\newcommand{\mean}[1]{\ensuremath{\left\langle #1 \right\rangle}}

\newcommand{\Trace}[1]{\ensuremath{\mathrm{Tr}\left [#1 \right]}}

\newcommand{\bra}[1]{\ensuremath{\left\langle #1 \right|}} 
\newcommand{\ket}[1]{\ensuremath{\left| #1 \right\rangle}}
 
\newcommand{\ketbra}[2]{\ensuremath{\left|#1\right\rangle \left\langle#2\right|}}

\newcommand{\ad}{\ensuremath{a^\dagger}}

\newcommand{\nl}{\nonumber \\}

\newcommand{\sigmaz}{\ensuremath{\sigma^{\textrm{z}}}}

\newcommand{\sigmap}{\ensuremath{\sigma^{\textrm{+}}}}
\newcommand{\sigmam}{\ensuremath{\sigma^{\textrm{--}}}}

\newcommand{\tcr}{\textcolor{red}}

\begin{document}

\title{Quantum trajectories, interference, and state localisation in dephasing assisted quantum transport}
\author{Kiran E. Khosla$^1$, Ardalan Armin$^2$, M. S. Kim$^{1,3}$}
\affiliation{$^1$ QOLS, Blackett Laboratory, Imperial College London, London SW7 2AZ, United Kingdom,}
\affiliation{$^2$ Sustainable Advanced Materials (S$\hat{e}$r SAM), Department of Physics, Swansea University, Swansea, Wales, SA2 8PP, U.K.}
\affiliation{$^3$ Korea Institute of Advanced Study, Seoul 02455, South Korea}

\begin{abstract}
Dephased quantum transport of excitations occurs when energetic fluctuations in a system are sufficient to suppress the built-up of coherent amplitudes. While this has been extensively studied in many different systems, a unified and comprehensive understanding of quantum assisted transport via on-site and coupling-induced dephasing processes is lacking. The aim of the present work is to present a simple and unified understanding of the role of these two key dephasing processes in dephasing assisted transport. Our work explicitly links continuous dephasing to classical and quantum transitions. We present a natural quantum trajectories explanation of how different coupling and dephasing terms alter the diffusion rate of excitations and how this is impacted by the onset of Anderson localized eigenstates. Our results provide insight in understanding quantum transport in molecular semiconductors, artificial lattices and quantum features of excitonic solids.
\end{abstract}
\maketitle

\section{Introduction}

Quantum transport studies the transmission of an initial quantum excitation \cite{kempe_quantum_2003,venegas-andraca_quantum_2012,aharonov_quantum_1993,chen_directional_2020,hagenmuller_cavity-enhanced_2017,whitfield_quantum_2010}, typically on a site- or lattice-based model. It has been extensively studied in a wide context of applications including field theoretic models \cite{haken_exactly_1973}, general lattice models \cite{caruso_universally_2014}, and charge generation in molecular solids \cite{bakulin_role_2012} to name a few. The presence of dephasing terms (i.e. non-unitary evolution) and even static disorder have been shown to increase the efficiency of quantum transport processes \cite{hagenmuller_cavity-enhanced_2017,hood_entropy_2016,coehoorn_effects_2012,plenio_dephasing-assisted_2008,ringsmuth_multiscale_2012}. This so-called Dephasing Assisted Transport has been suggested as possible explanation for the efficiently of photosynthesis \cite{caruso_highly_2009,caruso_coherent_2012} and organic photovoltaics \cite{bittner_noise-induced_2014}. Dephasing assisted transport is an example of stochastic resonance \cite{gammaitoni_stochastic_1998} where random fluctuations can amplify an otherwise low probability process. It has been well studied both experimentally \cite{broome_discrete_2010,biggerstaff_enhancing_2016,wang_efficient_2018,jiao_two-dimensional_2021,owens_two-photon_2011} and theoretically \cite{caruso_universally_2014,dalla_pozza_quantum_2020,rebentrost_quantum_2018,siloi_noisy_2017}, including the exploring the effects of non-Markovian fluctuations in the dephasing terms \cite{uchiyama_environmental_2018}.

Although dephasing assisted transport has been studied extensively in the past, a simple yet comprehensive understanding of different dephsing mechanisms across a wide range of parameter regimes is lacking. Filling this gap is the aim of the present work. We begin by introducing a simple 1D spin chain Hamiltonian, and qualitatively analyse how dephasing influences site-to-site transition probabilities. We then show how (i) on-site, and (ii) coupling-induced dephasing can both assist transport albeit via different mechanisms. For each mechanism we show how considering quantum trajectories \cite{wiseman_quantum_1996,daley_quantum_2014} give an intuitive picture of the transport process properties, and the qualitative behaviour of simulations can be readily interpreted. Furthermore, we describe the role of delocalised and localised eigenstates and how they may effect transport.


\section{Transport model}

\begin{figure}
	\includegraphics[width=.9\columnwidth]{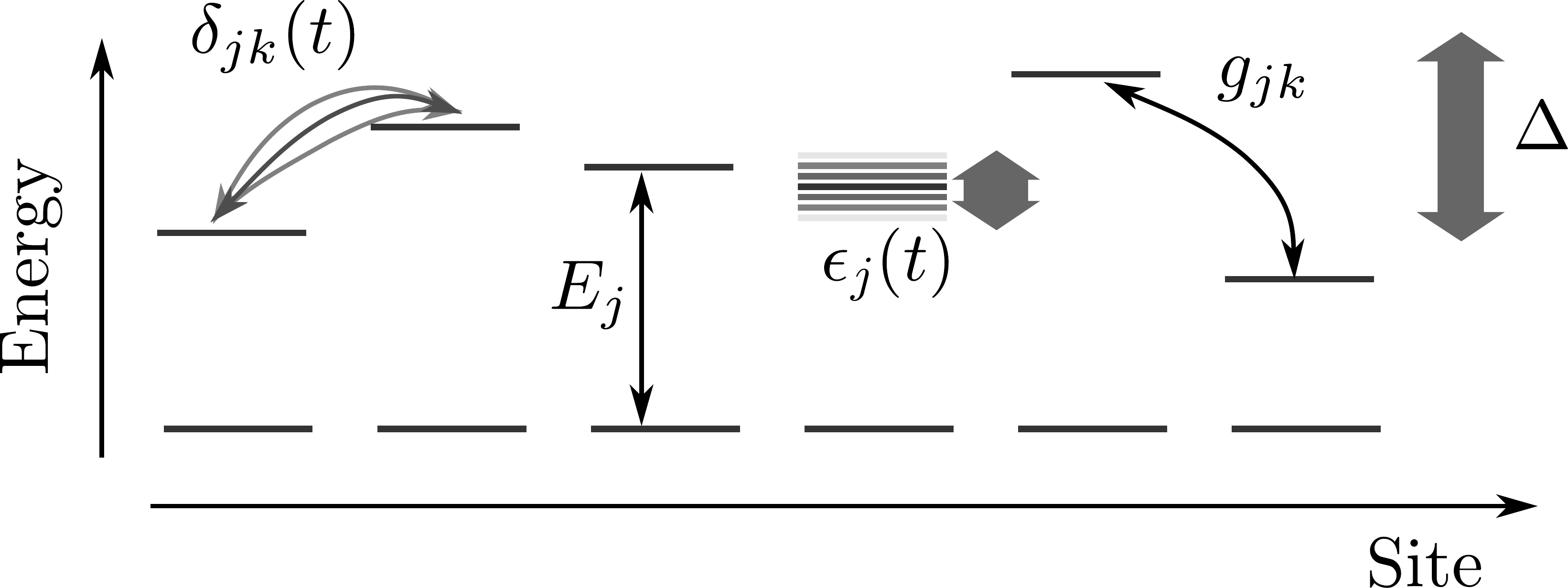}
	\caption{\label{fig:schmetaic} State diagram showing the coupling and dephasing terms. The on-site energy ($E_j$) and nearest neighbour couplings ($g$) are time-independent, with the disordered $E_j$ having characteristic spread of $\Delta$. On site energy and couplings fluctuate with independent $\epsilon_j(t)$ and $\delta_{jk}(t)$ respectively.}
\end{figure}

In the following we will consider a simple 1D site-based model. While extending the model system to higher dimensions is only slightly more technical, it is no more illuminating for an intuitive interpenetration. Each site $j$ in the $N$ site Hamiltonian has on-site energy $E_j$ (taking $\hbar = 1$), and is coupled to it's nearest neighbours via particle conserving transitions,
\begin{eqnarray}
H = \sum_i E_i \sigmaz_i + \sum_{<j,k>} g_{jk}( \sigmam_j\sigmap_k + \sigmam_k \sigmap_j) 
\label{eq:H}
\end{eqnarray}
where $\sigmaz$ is the usual Pauli-Z operator, $\sigmam$ ($\sigmap$) is the lowering (raising) operator, and $g_{jk}$ is the coupling rate between (neighbouring) sites $j$ and $k$. We introduce dephasing as a stochastic time dependent addition to the on-site energy  $E_j \rightarrow E_j + \epsilon_j(t)$, or coupling rates $g_{jk} \rightarrow g_{jk} + \delta_{jk}(t)$, see Fig.~\ref{fig:schmetaic}. We will set constant coupling rates $g_{jk} = g$, but the fluctuations on differing sites remain independent. The time-independent on-site energies $E_j$ are taken to be constant, but drawn from a normal distribution with variance $\Delta^2$, thereby adding static disorder to the system. In a physical system the time dependence arises from a bath of quantum degrees of freedom, for example neighbouring molecules/atoms, stray electromagnetic fields, or vibrational degrees of freedom. Treating these effects as a classical stochastic energy term is equivalent to an implicit Born-Markov approximation in the quantum system-bath coupling \cite{gardiner_input_1985} (which introduces a ultraviolet cut off, below which time scale the master equation is not longer Markovian). While not addressed here, we note that coupling the on-site energy to a single field degree of freedom results in polariton modes and has also been shown to speed up quantum transport processes \cite{hagenmuller_cavity-enhanced_2017}, however this mechanism is quite different to dephased case. Incoherent coupling terms can arise from dipole or Jaynes–Cummings or dispersive coupling to a single field degree of freedom \cite{zeb_incoherent_2020,strashko_organic_2018}.

Since the Hamiltonian preserves the total particle number $[H,~\sum_j \sigmaz_j] = 0$, initial states with a single excitation can be understood as a single particle wave function (e.g. single first quantised particle) over a discrete lattice. We will therefore use shorthand notation $\ket{j} \equiv \ket{...0_{j-1}1_j0_{j+1}..}$ as a state with excitation at site $j$. For constant coupling $g$ and no disorder ($E_i = E_j$) the single particle wave function obeys a simple (discrete) wave equation with linear dispersion. In this case a localised initial state propagates with well defined velocity $v$.

\subsection{Rabi amplitude intuition}

In the following show how Eq.~\eqref{eq:H} can already give insight as to how dephasing can assist or suppress transport. For the moment we focus on on-site dephasing only. Consider a single hopping term in the Hamiltonian in the interaction picture with $H_0 = \sum_j [E_j + \epsilon_j(t) ]\sigmaz_j$,
\begin{eqnarray}
H_{\mathrm{I},jk} = g e^{-i\phi_{jk}(t)}\sigmam_j \sigmap_k + \mathrm{H.C.}.
\end{eqnarray}
where $\phi_{jk}(t) = \int^t [E_j - E_k +\epsilon_j(\tau) - \epsilon_k(\tau)]d\tau$. This term generates Rabi-like\footnote{They are exactly Rabi oscillations if $\phi_{jk}(t) = (E_j - E_k) t$, and all other sites are neglected} oscillations (with coupling $g$ and detuning $d\phi/dt$) between sites $j$ and $k$ --- henceforth referred to as Rabi oscillations. The effective time dependent driving frequency is $\omega(t) \approx d\phi/dt$, and gives an effective (instantaneous) square Rabi amplitude \cite{Sakurai:1167961} of
\begin{eqnarray}
A_{jk}^2 = \frac{g^2}{g^2 + [E_j - E_k + \epsilon_j(t) - \epsilon_k(t) ]^2}
\label{eq:RabiA}
\end{eqnarray}
(as can be seen by diagonalising the $jk$ hoping terms in the Hamiltonian). Note $\epsilon_j(t)-\epsilon_k(t)$ is a stochastic process and a density matrix formalism is required before considering averages. However, by considering characteristic values of the noise terms\footnote{For a white noise process the characteristic value has to be averaged over a short but finite time, i.e. $\epsilon \equiv |\int^{\tau}\epsilon_j(t)-\epsilon_k(t) dt|$ with $\tau \ll g^{-1}$.} $\epsilon \equiv |\epsilon_j(t)-\epsilon_k(t)|$, we can use Eq.~\eqref{eq:RabiA} to qualitatively identify regimes where the noise will increase or decrease transition probabilities. As $\Delta$ quantifies the on-site disorder, the characteristic value of $|E_j - E_k|$ is simply $\Delta$. As the fluctuations $\epsilon_j(t)$ are zero-mean, the effective detuning term in the denominator takes values about the range $\approx ( \Delta \pm \epsilon)^2$, and in the following we will only consider the $\pm$ values, rather than the distribution.

For strongly unequal on-site energy $\Delta \gg g$ (i.e. strong disorder), the Rabi amplitude, Eq.~\eqref{eq:RabiA}, increases as $\epsilon$ approaches $\Delta$ from below. In this case the transition probability is assisted by the presence of dephasing terms. However, when the dephasing term becomes larger than the characteristic on-site energy difference, $\epsilon \gtrsim \Delta$, the dephasing contribution itself dominates the denominator of Eq.~\eqref{eq:RabiA} --- in this case dephasing is expected to suppress transport. For no disorder, $E_j = E_k$, any dephasing reduces the Rabi amplitude, and is therefore expected to suppress transport. 

The role of dephasing can now be intuitively understood for the case of non-negligible disorder $\Delta > g$. Weak dephsasing $\epsilon < \Delta$ allows more resonant site-to-site transitions, assisting transport. When the dephasing becomes sufficiently large $\epsilon > \Delta$, its starts pushing the transitions further from resonance, thereby suppressing transport. For zero disorder $\Delta = 0$, all transitions are pushed off resonance and dephasing suppresses transport. Refs. \cite{caruso_universally_2014,derrico_quantum_2013,caruso_fast_2016} show there is a dephasing rate that optimises quantum transport, and is consistent with the simple Rabi amplitude example in Eq.~\eqref{eq:RabiA} (e.g. $\epsilon \approx \Delta$). 

The above discussion has focused solely on the Rabi amplitude, and has neglected physically important considerations, such as interference, the many-site Hilbert space, and long time-scale effects. For example, when $\Delta  > g$, a 1D lattice exhibits Anderson localisation in the low energy (i.e. single particle) eigenstates \cite{anderson_absence_1958}. In this regime, the Rabi amplitude between neighbouring sites can be relatively large $A_{jk}^2\approx \mathcal{O}(10^{-1})$, yet long range transitions are suppressed by interference and an initial excitation quantisation on site $j$ can never propagate. Any	 observed long range transport of such a particle must be dephasing induced (or at least dynamically driven) irrespective of the finite Rabi amplitude. These points can be intuitively understood by considering the role of interference (and how it is suppressed by dephasing) in quantum transport as addressed in the following section. 

Firstly, we will consider on-site dephasing. The dynamics generated by Eq.~\eqref{eq:H} is analytically solvable for white noise dephasing and resulting equation of motion is equivalent to a stochastic measurement model. Secondly, we will consider analytic solution for fluctuations in the coupling rates. While both regimes are rightfully referred to as dephasing, the underlying physical mechanisms, and associated quantum trajectories are very different.

\section{Dephasing assisted transport}

\begin{figure}
	\includegraphics[width=\columnwidth]{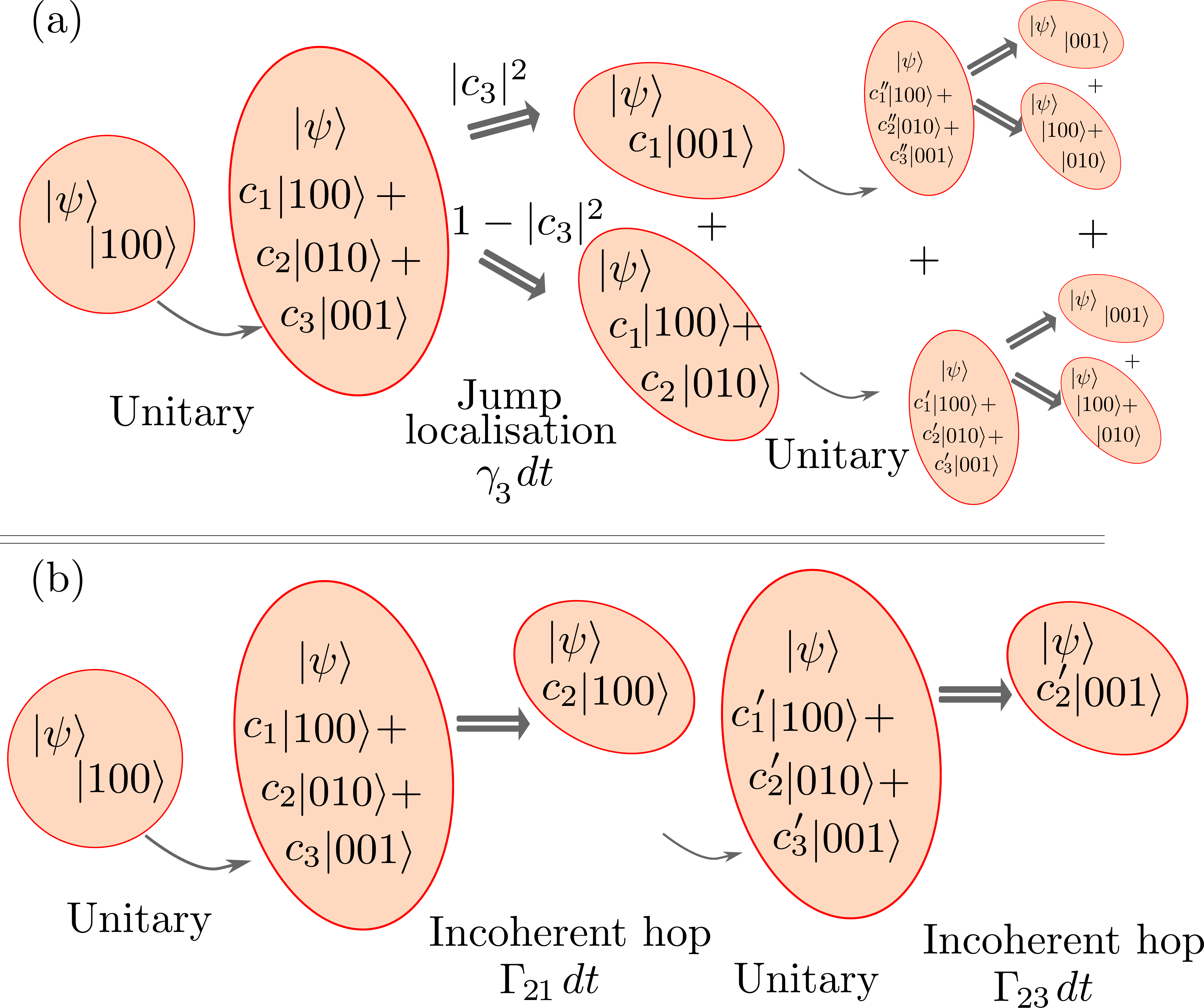}
	\caption{Qualitative difference between (a) dephasing assisted and (b) incoherent hopping transport mechanisms. (a) Fluctuations in the on-site energy are equivalent to a jump process (shown here on site $3$) with two possible wave function trajectories: (i) localise or (ii) exclude the excitation at a site with probabilities $|c_3|^2$ and $1-|c_3|^2$ respectively. For a fast jump rate $\gamma \gg g$, the probabilities $|c_i|^2$ (which are second order in $g t$) do not appreciably change, and the jump almost certainly localises the particle on it's initial site --- the onset of the Zeno effect. (b) Fluctuations in the site-to-site coupling rate cause a guaranteed incoherent hopping process that guarantees (state independent) localisation of the particle, shown here as hop two-to-one, followed by two-to-three. The incoherent hopping trajectories only have some probability to hop from site $j$ to $k$. In both cases a jump is random, and the phase relations in the post-jump quantum state are destroyed, but with different limiting behaviour.}
	\label{fig:jumps}
\end{figure}

In the case of Gaussian noise in the on-site energy, the master equation can be solved exactly and analytically (see appendix). The Lindblad master equation is

\begin{eqnarray}
\dot{\rho} = -i[H, \rho] + \sum_j\gamma_j \mathcal{L}[\sigmaz_j]\rho
\label{eq:weak_deph}
\end{eqnarray}
where $\mathcal{L}[A]\rho = A\rho A^\dagger - \frac12 A^\dagger A \rho - \frac12 \rho A^\dagger A$, and $\gamma$ is related to the spectral density the noise term, $\mathcal{E}[\epsilon_j(t)\epsilon_k(t')] = \gamma_j \delta_{jk}\delta(t-t')$. Henceforth we will take constant $\gamma_j = \gamma$. The Hamiltonian $H$ includes the coherent hopping $g$, and fixed on-site energy terms $E_j$. 

In Eq.~\eqref{eq:weak_deph} the jump operators are not the hopping terms $\sigmam_j \sigmap_k$, but rather jump localisation of excitations, $\sigmaz_j$. Nevertheless the $\sigmaz$ jumps can assist transport probabilities, even though there is no ``classical'' hopping process. Eq.~\eqref{eq:weak_deph} can be understood as a Poisson process \cite{wiseman_optimal_2005,wiseman_interpretation_1993,keys_poisson_2020} (and see appendix): with probability $2\gamma dt$, the $j$th site is projected onto an eigenstate of $\sigmaz_j$, Fig.~\ref{fig:jumps} (a), where the Born rule gives the probability of which eigenstate is projected. The two possible projections, as well as the Poisson process jumps are then incoherently averaged.

Under the free evolution of $H$, a single excitation at site $\ket{j}$ evolves into $U(t)\ket{j} = \sum_k c_{k} \ket{k}$, with probability amplitude $c_k$ to be found at site $\ket{k}$. A jump event at site $k$ happens with probability $2\gamma dt$ and has two probabilistic outcomes: (i) fully localises the excitation on site $k$ with total probability $2\gamma |c_{k}|^2 dt$; or (ii) fully exclude the possibility that the excitation is on site $k$ with total probability $2\gamma (1-|c_{k}|^2) dt$. This effectively resets the particle's location. However, whether this is assists or suppresses transport depends on the relative quantum and classical rates $g/\gamma$, and disorder $\Delta$. The fact that exclusion is as possible outcome is a consequence of the spin chain Hamiltonian. On-site energy de-phasing of a single (non-spin) particle on a lattice only has the localisation possibility as there is no notion of spin-down to project into.

The free evolution changes the probability amplitude of finding the excitation on some site (at a rate $g$), which in turn changes the probability at which the jump operator fully localises (or excludes) the excitation on a given site $\ket{k}$. The localisation process can increase the diffusion rate in an intuitive way: a localisation (or exclusion) event breaks phase relations that could otherwise lead to interference. The particle's location can therefore have a larger variance under this extra random process, depending on the disorder $\Delta$. For disorder free $E_j = E$, and constant coherent coupling $g$ the (effective single particle) wave equation causes destructive interference for trajectories where the particle returns to it's initial position. Breaking phase relations will suppresses transport and can be intuitively understood from (i) the Rabi amplitude argument above, or (ii) suppressing the destructive interferences that results in wave like propagation with a well defined velocity (e.g. $\Delta = 0$ in Fig.~\ref{fig:me_solves} (a)). We therefore focus on the disordered case ($\Delta > 0 $) in the following, differentiating between weak ($\Delta < g$) and strong ($\Delta > g$) disorder.

Even though the spin conserving interaction allows a single excitation to be treated as a single particle on a lattice, the $\sigmaz_j$ dephasing terms are are quite different to simple on-site dephasing terms for a single particle. The corresponding master equation for a single particle site-based model only has the possibility of localisation, i.e. there no possibility of exclusion. It is interesting that the existence of the larger Hilbert space of the spin chain, in contrast to a single particle on a lattice, has a material effect on allowed quantum trajectories, even though the system always remains in the single particle subspace.

Fig.~\ref{fig:me_solves} (a) and (b) show the evolution of the site dephased master equation, Eq.~\eqref{eq:weak_deph}, plotting the probability distribution over times. In each case, the initial state is a single excitation in the middle of a 201 site chain. The plots systematically covers all parameter regimes with rows corresponding to zero ($\Delta = 0$), weak ($\Delta = g/2$), intermediate ($\Delta = g$), and strong ($\Delta  > g$) disorder regimes. Fig.~\ref{fig:me_solves} (a) shows the unitary solution (with $\gamma = 0$), and columns in Fig.~\ref{fig:me_solves} correspond to weak ($\gamma = g/10$), intermediate ($\gamma = g$), and strong ($\gamma = 10g$) dephasing regimes. In the following we discuss the interpretation for on-site dephasing.

\subsection{Weak dephasing}

For weak dephsing $g \gg \gamma$, there are $g/\gamma \gg 1$ coherent transitions between jump localisation events. Hence on the localisaiton time scale $\gamma^{-1}$, the amplitudes $|c_k|^2$ change sufficiently fast, and can be approximated by their time averages (over $\gamma^{-1}$). This allows a coherent quantum transport process \cite{derrico_quantum_2013,biggerstaff_enhancing_2016} with an effective resetting of the excitation's location on the $\gamma^{-1}$ time scale. For weak disorder $0 < \Delta < g$, wave-like transport causes fast diffusion and the corresponding probability of finding he particle on any specific site, $|c_k|^2$ is low. In this case, for a given localisation event, the probability of exclusion is significantly higher than the probability of localisation. 

\begin{widetext}
	
	\begin{figure}
		\includegraphics[width=\textwidth]{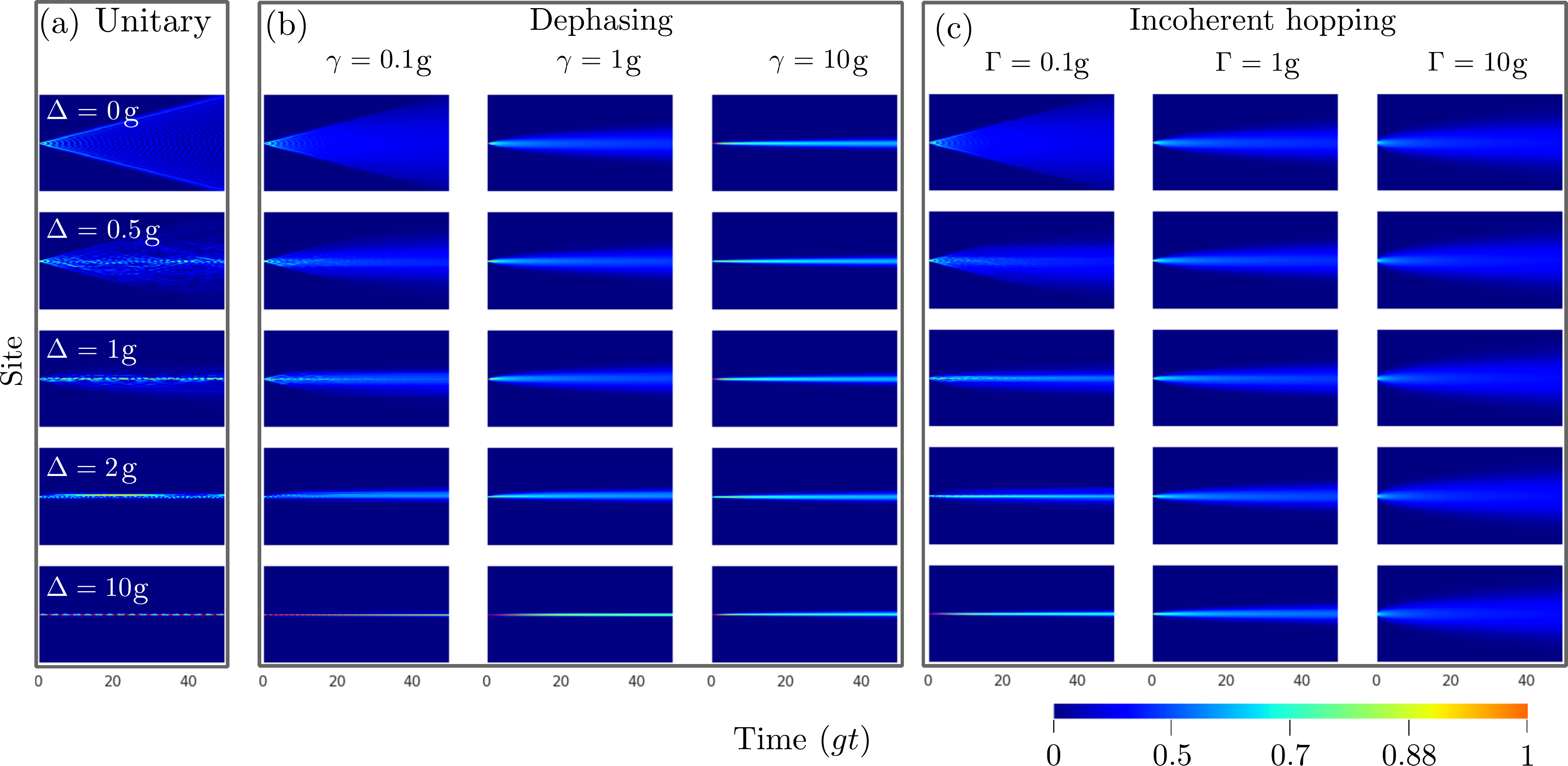}
		\caption{\label{fig:me_solves} Time evolution of probability distributions given a single excitation in the middle of a 201 site spin chains, showing different disorder (rows) and dephasing rates (columns). (a) Unitary evolution (no dephasing) showing the onset of Anderson localisation with increasing $\Delta$ (b) On site dephasing master equation, Eq.~\eqref{eq:weak_deph}. For increasing $\gamma$ the Zeno effect is clearly seen, becoming more apparent with increasing $\Delta$. (c) Incoherent hopping master equation, Eq.~\eqref{eq:strong_deph}. As $\Gamma$ increases the solutions closely resembles a classical random walk, and the disorder $\Delta$ has little effect.  In both master equation solutions, (b) and (c), there is a noticeable change from unitary wave-like propagation ($\Delta = 0$) with a clear velocity, to the approx approximate classical random walk, with propagation distance proportional to $\sqrt{t}$.}
	\end{figure}
	
\end{widetext}

Dephasing assisted transport becomes pronounced as site-to-site disorder becomes comparable to, or larger than the coupling rate, i.e. the Anderson localisation regime, $\Delta \approx g$ or $\Delta > g$, Fig.~\ref{fig:me_solves} (b). Without dephasing, an initial excitation is forever localised around it's initial state (e.g. see Fig.~\ref{fig:me_solves} (a) $\Delta = 10$). This can be understood from energetic \cite{anderson_absence_1958}, interference \cite{schwartz_transport_2007}, and active Hilbert space perspectives. When the $\sigmaz$ jump process is included, there is some chance for the excitation to fully jump to a neighbouring site. From here it remains Anderson localised, but now at a new site. This adds classical random walk-like dynamics to the particle, giving mobility to an otherwise localised excitation. However, as the probability for coherently hopping to new site is low (i.e. Anderson suppressed), the jump localisation has a higher chance to localise the particle on it's initial site. Hence, while there is an additional classical random walk happening, it is not simply given by the Poisson rate $\gamma$. This becomes particularity apparent in the strong dephasing limit.

\subsection{Strong dephasing} 
\label{sec:strong_deph}

As $\gamma$ becomes larger than the coherent hopping rate $g$, the jump localisation begins to suppress the transport probability. This could be qualitatively seen from the Rabi transition amplitude, but the mechanism is very simple in the jump localisation picture: it is the onset of the Zeno effect (see Fig.~\ref{fig:me_solves} (b) $\gamma = 10$). For strong dephasing, there are $\gamma/ g > 1$ incoherent jumps per coherent transition, and there is insufficient time for the coherent evolution to appreciably change the amplitude of the excitation to be found on a neighbouring site. This is a consequence of the probability of a coherent jump being a second order effect ($P_{\mathrm{coherent~hop}} \equiv |c_k(dt)| = g^2dt^2$ for $c_k(0) = 0$), while the Poisson jump process is a first order effect, $P_{\mathrm{jump}} =\gamma dt$. The onset of the Zeno effect is independent of any on-site disordered, but Anderson localisation additionally reduces the transport probability.

The scaling of the classical hopping rate can be approximated by considering the probability for a coherent jump over the $\gamma^{-1}$ time scale, and how this effects the probability of localisation in the Poisson process. Given an initial excitation fully localised on site 0, the probability at time $t \ll g$ of a coherent transition to site 1 is $|c_1(t)|^2 = g^2 t^2$, and the probability per unit time of a jump localisation on site 1 is therefore $g^2 t^2 \times \gamma$. On the time scale of $g^{-1}$ the average probability per unit time that the Poisson process localises the state on site 1 is $P_{0\rightarrow 1} = \frac{1}{\gamma^{-1}} \int^{1/\gamma} g^2 t^2 \gamma dt \propto g^2/\gamma$. Hence in the $\gamma/g \gg 1$ regime, increasing $\gamma$ in fact decreases the effective classical Poisson hopping rate. Nevertheless, the existence of a fix hopping probability per-unit time means strong dephasing asymptotically becomes a classical hopping process, albeit with asymptomatically vanishing hopping rate. This can be see in the $\gamma = 10$ column of Fig.~\ref{fig:me_solves} (b), where there is some classical-like, albeit very slow diffusion, and this is further suppressed by Anderson localisation (e.g. compare $\Delta = 0$ and $\Delta = 10$ in Fig.~\ref{fig:me_solves} (b)).

Eq.~\eqref{eq:weak_deph} and the results in Fig.~\ref{fig:me_solves} (b) can now be readily understood in the the jump localisation picture. Given some disorder, weak dephasing terms can increases the transport probability by probabilistically localising the excitation to new sites. While this is an effective Poisson walk, the underlying probabilities of the Poisson walk depend on the probability amplitudes, and therefore the coherent evolution. This has an especially large effect for highly disordered systems where Anderson localisation would otherwise prevent transport. As the dephasing rate increases the onset of the Zeno effect begins to suppress the effective classical hopping rate.

\begin{widetext}	
	
	\begin{figure}
		\includegraphics[width=\textwidth]{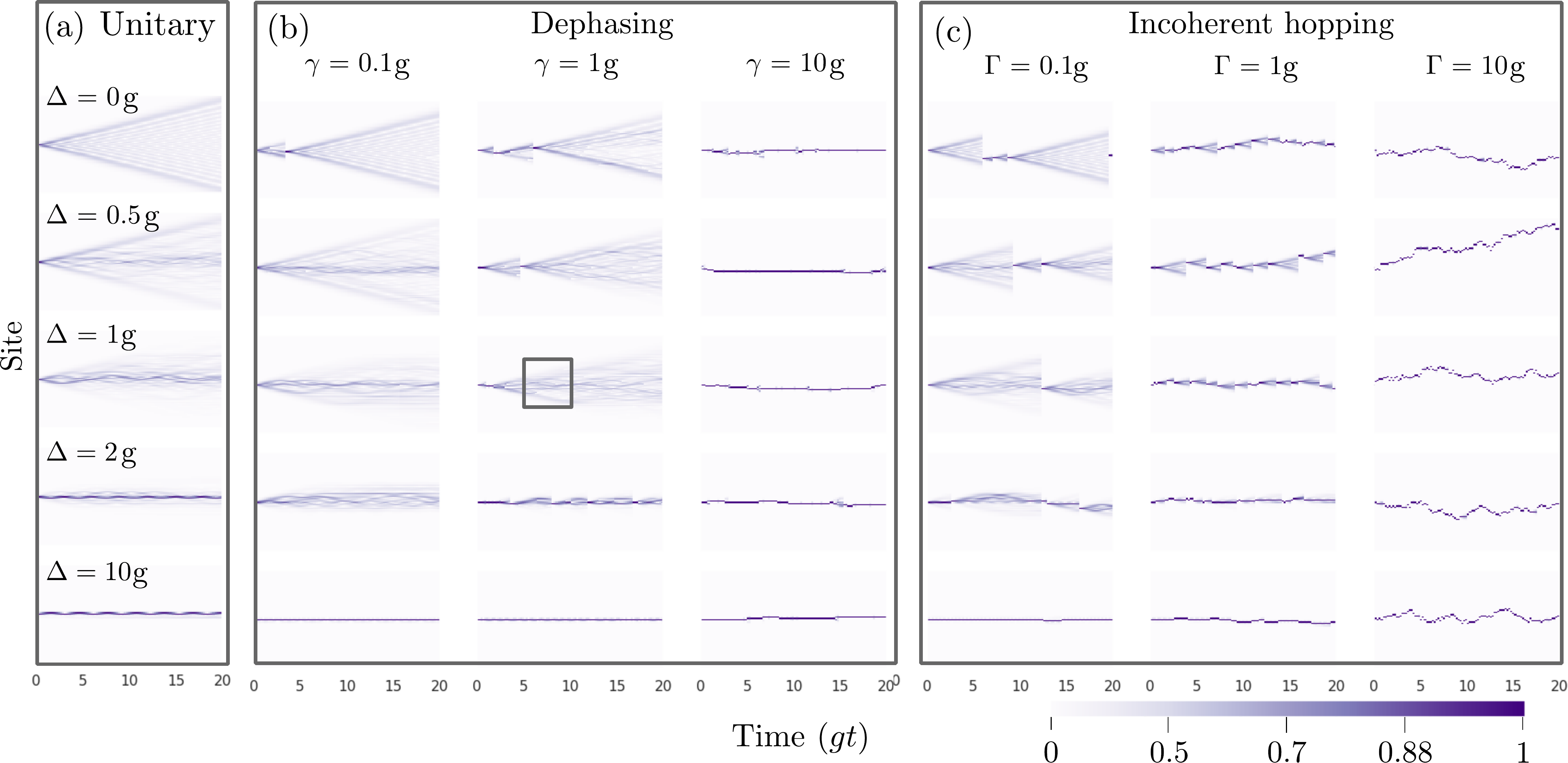}
		\caption{\label{fig:traj_solves} Single quantum trajectories a single excitation in the middle of a 81 site spin chains, showing the disorder (rows) and dephasing rates (columns). (a) Unitary evolution (no dephasing terms). (b) Quantum trajectory corresponding to on-site dephasing, Eq.~\eqref{eq:weak_deph}, showing the clear Zeno effect for $\gamma = 10g$, however the classicla random walk is more obvious, particular for $\Delta = 0$. Localisation events show clear resetting of the wave-like propagation. Jump exclusion events are not as obvious here (see Fig.~\ref{fig:traj_zoom} for examples of exclusion events). (c) Incoherent hopping quantum trajectories of Eq.~\eqref{eq:strong_deph}. For a given $\gamma = \Gamma$, the incoherent hopping localisation event appear far more frequent as there is no possibility for exclusion. The limiting behaviour to a classical random walk is far more clear with increasing $\Gamma$, and in this limit $\Delta$ plays almost no role.}
	\end{figure}
	
\end{widetext}

\section{Incoherent transport}

Incoherent transport can also arise from classical fluctuations ($\delta_{jk}(t)$ in Fig.~\ref{fig:schmetaic}) in the coupling term. These fluctuations also result in loss of phase coherence, and in effectively the same basis. For clarity we will refer to this effect as incoherent transport instead of dephased transport. Averaging over fluctuations in the coupling terms result in the following non-perturbative master equation (see appendix),

\begin{eqnarray}
\dot{\rho} &=& -i[H,\rho] + \sum_{<jk>} \Gamma_{jk}\mathcal{D}[\sigmam_j\sigmap_k]\rho.
\label{eq:strong_deph}
\end{eqnarray}
We will consider the case of fixed $\Gamma_{jk} = \Gamma$. The the Lindblad terms admit a more obvious Poisson process interpretation: with probability $2\Gamma dt$ a particle incoherently hops from site $j$ to $k$ (see appendix for details). The Poisson hopping mechanism is far more intuitive than the localisation mechanism for the on-site dephasing case, and it nicely interpolates between classical and quantum transport processes. A similar incoherent hopping term can arise from coupling the spins to a field degree of freedom which is then (perturbatively) adiabatically eliminated~\cite{zeb_incoherent_2020,strashko_organic_2018}. When a particle hops from form $j$ to $k$, any phase relation between the all other amplitudes and $c_j$ vanish, Fig. \ref{fig:jumps} (b). Again, this jump operator can again assist or suppress transport depending on the other rates in Hamiltonian. However, unlike the dephasing case where the Zeno effect completely suppresses transport to almost zero, ``suppressed transport'' here means a classical Poisson hopping process. Numerical solutions of the incoherent hopping master equation, Eq.~\eqref{eq:strong_deph} are shown in Fig.~\ref{fig:me_solves} (c), covering the same regimes as site dephasing master equation.

In contrast to the dephasing regime, there is no state-dependent probability of localisation or exclusion, Fig.~\ref{fig:jumps}. Here the random hopping event always moves the particle from site $j$ to $k$, entirely localising it at site $k$, and effectively resetting the initial state. While there is no state dependent probability for the jump event happening, the relative contribution of this event to the ensemble average does depend on the state. Fig.~\ref{fig:me_solves} (c) plots the probability distribution (over a 201 site chain) over time for the incoherent hopping master equation, Eq.~\eqref{eq:strong_deph} with the same parameters as Fig.~\ref{fig:me_solves} (c), and shows different limiting behaviour in $\Gamma$.

\subsection{Weak incoherent coupling}

When the incoherent process is slow compared to the coherent coupling $\Gamma < g$, there is sufficient time to allow the probability amplitudes to coherently evolve. If the on-site energy is only weakly disordered $\Delta < g$ (i.e. no eigenstate localisation), the slow classical hopping rates allow probability amplitudes to appreciably build up giving fast wave-like quantum transport over the short $t< \Gamma^{-1}$ time scale, with relocalisation events probabilistically happening over the $t > \Gamma^{-1}$ time scale. The incoherent hops break the destructive interference necessary that gives the fast wave-like prorogation, and therefore are expected to slow down the overall transport. 

Similar to on-site dephasing, weak incoherent hopping is effected by Anderson localisation, albeit not as much. Under incoherent hopping, a jump operator $\sigmam_j \sigmap_k$ acting on an excitation localised at site $j$ will localise it on site $k$, irrespective of any coherent amplitudes $c_j$, $c_k$. However, any free evolution between hopping events does not significantly contribute to the overall transport (in the Anderson case), see $\Gamma = 0.1$ in Fig.~\ref{fig:me_solves} (c). Hence in the Anderson regime, the only appreciable transport mechanism is via the classical hopping process, and the effective diffusion rate is expected to asymptote to classical Poisson diffusion with rate $\Gamma$. This is in contrast to the weak on-site dephasing case, where transport rate is asymptotically suppresses as the localised eigenstates result in (asymptotically) no evolution of probability amplitudes, and therefore no possibility for the particle to become localised to any other state.

\subsection{Strong incoherent coupling}
Intuitively, the strong incoherent coupling regime, $\Gamma \gtrsim g$, asymptomatically approaches a classical Poisson random walk, irrespective of the disorder $\Delta$, see $\Gamma = 10$ in Fig.~\ref{fig:me_solves} (c). This is in contrast to DAT where the Zeno effect prevents any and all transport for strong dephasing. For $\Gamma \gg g$, there are $\Gamma / g \gg 1$ classical Poisson jumps for every single coherent transition. The small coherent coupling will always increase the transport probability, but the coherent effect effect becomes negligible as the number of Poisson jumps per coherent jump increase. As the coherent change in probabilities are at least second order in time (when starting with a fully localised initial condition), while Poisson jumps are first order, the classical diffusion process quickly dominates for $\Gamma > g$.

For very strong coherent coupling, disorder in the on-site energy plays very little role. The classical hopping term does not require any energy conservation --- the jump is equally likely for equal on-site energies, and completely different on-site energies. This is a simple consequence of the $\delta_{jk}(t)$ term taken to be a (spectrally flat) white noise process that can supply or absorb any amount of energy.

\section{Quantum Trajectories}
\label{sec:quantum_trajectories}

\begin{figure}
	\includegraphics[width=\columnwidth]{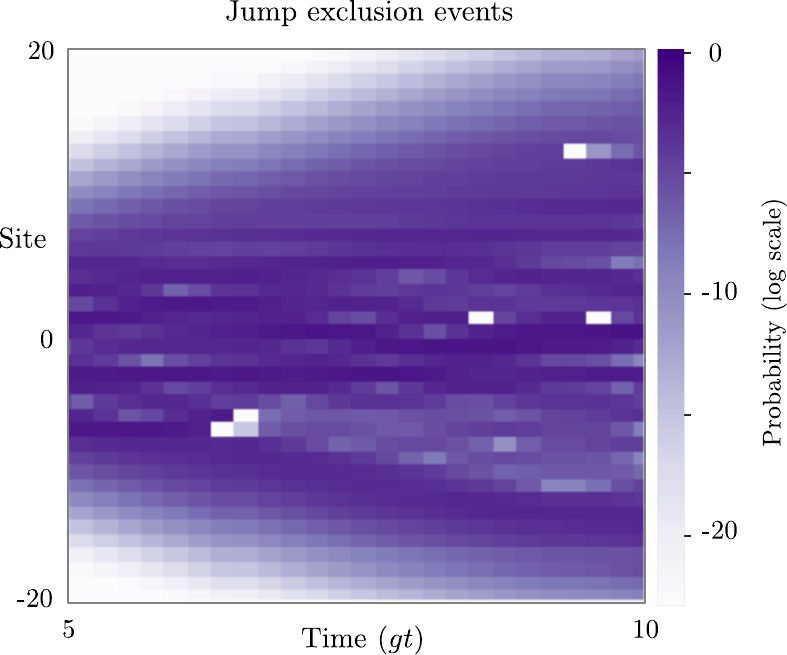}
	\caption{\label{fig:traj_zoom} Exclusion process (white holes) in the quantum trajectory of the probability distribution on-site dephasing. Zoomed in plot from Fig.~\ref{fig:traj_solves} (b), $\Delta = \gamma = 1$ . Such exclusion jump process are a consequence of the spin chain model, and are not present for a single particle random walk with site dephasing.}
\end{figure}

\begin{figure}
	\includegraphics[width=\columnwidth]{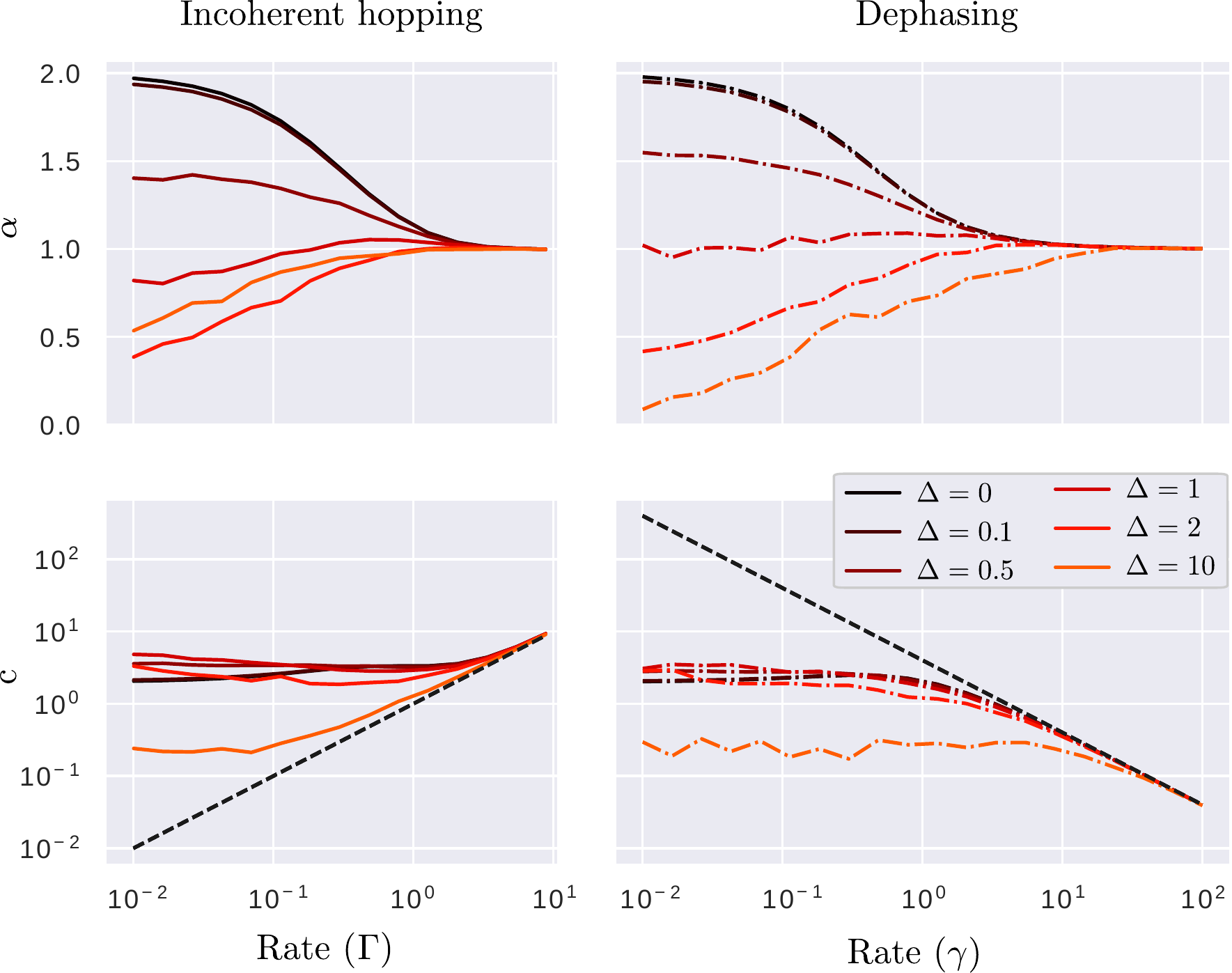}
	\caption{\label{fig:fits} Fitting parameters for the mean-square distance travelled $\mathcal{D}_{\mathrm{MS}}(t) = c t^\alpha$ showing the limiting classical random walk $\alpha = 1$. For a given $\gamma, \Gamma$, larger disorder (lighter colours) consistently has the slower (in $c$ and $\alpha$) diffusion. For weak disorder $\Delta \leq 1/2$, increasing $\Gamma$ or $\gamma$ decreases $\alpha$ with little change to $c$, thereby suppressing transport. Dephasing can be said to increase the diffusion for $\Delta > 1$ as $\alpha$ increases with increasing $\Gamma, \gamma$. However, increasing $\gamma$ in this regime reduces the coefficient $c$, and the notion of optimal transport becomes a balance between increasing $\alpha$ and decreasing $c$. The value of $\gamma$ where these effects balance must therefore depend on $\Delta$ and the system size. Dashed lines show $c = \Gamma$ and $c = 1/\gamma$. Rates, $\Gamma, \gamma$ and disorder $\Delta$ are quoted in units of $g$.}
\end{figure}

We numerically simulate the quantum trajectories of the on-site dephasing and incoherent transport. Weak and strong dephasing regimes (as in Fig.~\ref{fig:me_solves}) are covered to show example processes discussed in previous sections. Each trajectory represents a pure state unravelling of the corresponding master equation. In each case we consider a single excitation initially located in the centre of a 81 site chain, and simulate a trajectory for $gt = 20$. 

Fig.~\ref{fig:traj_solves} plots example quantum trajectories highlighting the effective difference between unitary (Fig.~\ref{fig:traj_solves} (a)) dephasing (Fig.~\ref{fig:traj_solves} (b)) and incoherent hopping (Fig.~\ref{fig:traj_solves} (c)). For incoherent hopping, every jump event completely re-localises the particle as the $\sigmam_j \sigmap_k$ term annihilates all but the $\ket{j}$ state, which jumps to $\sigmam_j \sigmap_k\ket{j} = \ket{k}$ . In contrast, the dephasing trajectories have two possible outcomes for every hopping event: localise or exclude. If the single particle wave function spreads sufficiently far between hopping events, the amplitude for any single site is relatively low, and the hopping events more often exclude rather than localise the particle. Fig.~\ref{fig:traj_zoom} (log colour scale) shows the exclusion processes for the $\gamma = \Delta = 1$ quantum trajectory of Fig.~\ref{fig:traj_solves}.

From the quantum trajectories, Fig.~\ref{fig:traj_solves}, it is clear the onset of Anderson localisation has little effect for strong incoherent hopping ($\Gamma = 10$) which well approximates a classical hopping process irrespective of $\Delta$. In contrast, Anderson localisation and the Zeno combine to reduce the diffusion rate for site dephasing. There is a noticeable, but highly suppressed classical random walk process for $\gamma = \Delta = 10$, which is not apparent for $\Delta = 10$, $\gamma = 0, 0.1$. The asymmetric ripples in the unitary case are due to a single instance of the randomly chosen on-site energy terms $E_j$ from a zero mean normal distribution and variance $\Delta^2$ (adding a non-zero mean to each $E_j$ does not change the dynamics).

Finally we analyse the diffusion rate and coefficients as a function of dephasing and disorder. For a classical hopping process from an initially localised point, the mean square distance travelled (denoted by $\mathcal{D}_{MS}$) increases as $\mathcal{D}_{MS}(t) = Dt$ with diffusion constant $D$. For wave-like propagation the increase scales as $\mathcal{D}_{MS}(t) = v^2t^2$ with velocity $v$. Fig.~\ref{fig:fits} plots power and coefficients of mean square diffusion $\mathcal{D}_{MS}(t)= c t^\alpha$ as fitted to numerical simulations. Curves are averaged over one hundred instances of random on-site disorders $\{E_j\}$. For both master equations the diffusion approaches classical hopping with $\alpha = 1$, but with oppositely scaling coefficients. For incoherent hopping, the diffusion rate approaches $\Gamma$ as expected. The dephasing case also approaches a classical random walk, but the diffusion rate becomes proportional to $1/\gamma$, as shown in section~\ref{sec:strong_deph}.

\section{Conclusion}

We have analysed two types of dephasing: on-site [Eq.~\eqref{eq:weak_deph}], and site-to-site [Eq.~\eqref{eq:strong_deph}], for a simple spin chain model. We systematically coved weak, strong, ordered and disordered parameter regimes for both types of dephasing. Master equations were derived analytically for the simple case of white noise dephasing, and the qualitative difference between the jump processes were discussed. Given strong dephasing, both master equations results in an effective classical Poisson random walk, but with opposite limiting Poisson rates: $\propto \Gamma$ for incoherent hopping, and $\propto 1/\gamma$ for on-site dephasing. 

In the case of no disorder, we noted that a single excitation in our model obeys a wave-equation on a discrete lattice (e.g. $\Delta = 0$ row in  Fig.~\ref{fig:me_solves} (a)), and any dephasing reduces the diffusion rate $\mathcal{D}_{MS}(t)$. However, when disorder becomes comparable to, or greater than the coherent coupling rate $\Delta \gtrsim g$, dephasing can increase the diffusion rate, e.g. see $\Delta \geq 1$ in Fig.~\ref{fig:me_solves}) with $\gamma = 0.1, 1, 10$, or $\Gamma = 0.1, 1, 10$.

\section{Acknowledgements}
This work is supported by the UK Hub in Quantum Computing and Simulation, part of the UK National Quantum Technologies Programme with funding from UKRI EPSRC grant EP/T001062/1, and the Korea Institute of Science and Technology (KIST) Open Research Program. AA is a S$\mathrm{\hat{e}}$r Cymru II Rising Star Fellow and acknowledges support from Swansea University strategic initiative in Sustainable Advanced Materials (S$\mathrm{\hat{e}}$r-SAM) funded by the S$\mathrm{\hat{e}}$r Cymru II Program through the European Regional Development Fund, and Welsh European Funding Office.


\begin{thebibliography}{41}%
	\makeatletter
	\providecommand \@ifxundefined [1]{%
		\@ifx{#1\undefined}
	}%
	\providecommand \@ifnum [1]{%
		\ifnum #1\expandafter \@firstoftwo
		\else \expandafter \@secondoftwo
		\fi
	}%
	\providecommand \@ifx [1]{%
		\ifx #1\expandafter \@firstoftwo
		\else \expandafter \@secondoftwo
		\fi
	}%
	\providecommand \natexlab [1]{#1}%
	\providecommand \enquote  [1]{``#1''}%
	\providecommand \bibnamefont  [1]{#1}%
	\providecommand \bibfnamefont [1]{#1}%
	\providecommand \citenamefont [1]{#1}%
	\providecommand \href@noop [0]{\@secondoftwo}%
	\providecommand \href [0]{\begingroup \@sanitize@url \@href}%
	\providecommand \@href[1]{\@@startlink{#1}\@@href}%
	\providecommand \@@href[1]{\endgroup#1\@@endlink}%
	\providecommand \@sanitize@url [0]{\catcode `\\12\catcode `\$12\catcode
		`\&12\catcode `\#12\catcode `\^12\catcode `\_12\catcode `\%12\relax}%
	\providecommand \@@startlink[1]{}%
	\providecommand \@@endlink[0]{}%
	\providecommand \url  [0]{\begingroup\@sanitize@url \@url }%
	\providecommand \@url [1]{\endgroup\@href {#1}{\urlprefix }}%
	\providecommand \urlprefix  [0]{URL }%
	\providecommand \Eprint [0]{\href }%
	\providecommand \doibase [0]{http://dx.doi.org/}%
	\providecommand \selectlanguage [0]{\@gobble}%
	\providecommand \bibinfo  [0]{\@secondoftwo}%
	\providecommand \bibfield  [0]{\@secondoftwo}%
	\providecommand \translation [1]{[#1]}%
	\providecommand \BibitemOpen [0]{}%
	\providecommand \bibitemStop [0]{}%
	\providecommand \bibitemNoStop [0]{.\EOS\space}%
	\providecommand \EOS [0]{\spacefactor3000\relax}%
	\providecommand \BibitemShut  [1]{\csname bibitem#1\endcsname}%
	\let\auto@bib@innerbib\@empty
	\bibitem [{\citenamefont {Kempe}(2003)}]{kempe_quantum_2003}%
	\BibitemOpen
	\bibfield  {author} {\bibinfo {author} {\bibfnamefont {J.}~\bibnamefont
			{Kempe}},\ }\href {\doibase 10.1080/00107151031000110776} {\bibfield
		{journal} {\bibinfo  {journal} {Contemporary Physics}\ }\textbf {\bibinfo
			{volume} {44}},\ \bibinfo {pages} {307} (\bibinfo {year} {2003})},\ \bibinfo
	{note} {publisher: Taylor \& Francis \_eprint:
		https://doi.org/10.1080/00107151031000110776}\BibitemShut {NoStop}%
	\bibitem [{\citenamefont
		{Venegas-Andraca}(2012)}]{venegas-andraca_quantum_2012}%
	\BibitemOpen
	\bibfield  {author} {\bibinfo {author} {\bibfnamefont {S.~E.}\ \bibnamefont
			{Venegas-Andraca}},\ }\href {\doibase 10.1007/s11128-012-0432-5} {\bibfield
		{journal} {\bibinfo  {journal} {Quantum Inf Process}\ }\textbf {\bibinfo
			{volume} {11}},\ \bibinfo {pages} {1015} (\bibinfo {year}
		{2012})}\BibitemShut {NoStop}%
	\bibitem [{\citenamefont {Aharonov}\ \emph {et~al.}(1993)\citenamefont
		{Aharonov}, \citenamefont {Davidovich},\ and\ \citenamefont
		{Zagury}}]{aharonov_quantum_1993}%
	\BibitemOpen
	\bibfield  {author} {\bibinfo {author} {\bibfnamefont {Y.}~\bibnamefont
			{Aharonov}}, \bibinfo {author} {\bibfnamefont {L.}~\bibnamefont
			{Davidovich}}, \ and\ \bibinfo {author} {\bibfnamefont {N.}~\bibnamefont
			{Zagury}},\ }\href {\doibase 10.1103/PhysRevA.48.1687} {\bibfield  {journal}
		{\bibinfo  {journal} {Phys. Rev. A}\ }\textbf {\bibinfo {volume} {48}},\
		\bibinfo {pages} {1687} (\bibinfo {year} {1993})},\ \bibinfo {note}
	{publisher: American Physical Society}\BibitemShut {NoStop}%
	\bibitem [{\citenamefont {Chen}\ \emph {et~al.}(2020)\citenamefont {Chen},
		\citenamefont {Ma},\ and\ \citenamefont {Sun}}]{chen_directional_2020}%
	\BibitemOpen
	\bibfield  {author} {\bibinfo {author} {\bibfnamefont {J.-F.}\ \bibnamefont
			{Chen}}, \bibinfo {author} {\bibfnamefont {Y.-H.}\ \bibnamefont {Ma}}, \ and\
		\bibinfo {author} {\bibfnamefont {C.-P.}\ \bibnamefont {Sun}},\ }\href
	{\doibase 10.1007/s11467-019-0944-x} {\bibfield  {journal} {\bibinfo
			{journal} {Front. Phys.}\ }\textbf {\bibinfo {volume} {15}},\ \bibinfo
		{pages} {21602} (\bibinfo {year} {2020})}\BibitemShut {NoStop}%
	\bibitem [{\citenamefont {Hagenmüller}\ \emph {et~al.}(2017)\citenamefont
		{Hagenmüller}, \citenamefont {Schachenmayer}, \citenamefont {Schütz},
		\citenamefont {Genes},\ and\ \citenamefont
		{Pupillo}}]{hagenmuller_cavity-enhanced_2017}%
	\BibitemOpen
	\bibfield  {author} {\bibinfo {author} {\bibfnamefont {D.}~\bibnamefont
			{Hagenmüller}}, \bibinfo {author} {\bibfnamefont {J.}~\bibnamefont
			{Schachenmayer}}, \bibinfo {author} {\bibfnamefont {S.}~\bibnamefont
			{Schütz}}, \bibinfo {author} {\bibfnamefont {C.}~\bibnamefont {Genes}}, \
		and\ \bibinfo {author} {\bibfnamefont {G.}~\bibnamefont {Pupillo}},\ }\href
	{\doibase 10.1103/PhysRevLett.119.223601} {\bibfield  {journal} {\bibinfo
			{journal} {Phys. Rev. Lett.}\ }\textbf {\bibinfo {volume} {119}},\ \bibinfo
		{pages} {223601} (\bibinfo {year} {2017})}\BibitemShut {NoStop}%
	\bibitem [{\citenamefont {Whitfield}\ \emph {et~al.}(2010)\citenamefont
		{Whitfield}, \citenamefont {Rodríguez-Rosario},\ and\ \citenamefont
		{Aspuru-Guzik}}]{whitfield_quantum_2010}%
	\BibitemOpen
	\bibfield  {author} {\bibinfo {author} {\bibfnamefont {J.~D.}\ \bibnamefont
			{Whitfield}}, \bibinfo {author} {\bibfnamefont {C.~A.}\ \bibnamefont
			{Rodríguez-Rosario}}, \ and\ \bibinfo {author} {\bibfnamefont
			{A.}~\bibnamefont {Aspuru-Guzik}},\ }\href {\doibase
		10.1103/PhysRevA.81.022323} {\bibfield  {journal} {\bibinfo  {journal} {Phys.
				Rev. A}\ }\textbf {\bibinfo {volume} {81}},\ \bibinfo {pages} {022323}
		(\bibinfo {year} {2010})},\ \bibinfo {note} {publisher: American Physical
		Society}\BibitemShut {NoStop}%
	\bibitem [{\citenamefont {Haken}\ and\ \citenamefont
		{Strobl}(1973)}]{haken_exactly_1973}%
	\BibitemOpen
	\bibfield  {author} {\bibinfo {author} {\bibfnamefont {H.}~\bibnamefont
			{Haken}}\ and\ \bibinfo {author} {\bibfnamefont {G.}~\bibnamefont {Strobl}},\
	}\href {\doibase 10.1007/BF01399723} {\bibfield  {journal} {\bibinfo
			{journal} {Z. Physik}\ }\textbf {\bibinfo {volume} {262}},\ \bibinfo {pages}
		{135} (\bibinfo {year} {1973})}\BibitemShut {NoStop}%
	\bibitem [{\citenamefont {Caruso}(2014)}]{caruso_universally_2014}%
	\BibitemOpen
	\bibfield  {author} {\bibinfo {author} {\bibfnamefont {F.}~\bibnamefont
			{Caruso}},\ }\href {\doibase 10.1088/1367-2630/16/5/055015} {\bibfield
		{journal} {\bibinfo  {journal} {New J. Phys.}\ }\textbf {\bibinfo {volume}
			{16}},\ \bibinfo {pages} {055015} (\bibinfo {year} {2014})},\ \bibinfo {note}
	{publisher: IOP Publishing}\BibitemShut {NoStop}%
	\bibitem [{\citenamefont {Bakulin}\ \emph {et~al.}(2012)\citenamefont
		{Bakulin}, \citenamefont {Rao}, \citenamefont {Pavelyev}, \citenamefont {van
			Loosdrecht}, \citenamefont {Pshenichnikov}, \citenamefont {Niedzialek},
		\citenamefont {Cornil}, \citenamefont {Beljonne},\ and\ \citenamefont
		{Friend}}]{bakulin_role_2012}%
	\BibitemOpen
	\bibfield  {author} {\bibinfo {author} {\bibfnamefont {A.~A.}\ \bibnamefont
			{Bakulin}}, \bibinfo {author} {\bibfnamefont {A.}~\bibnamefont {Rao}},
		\bibinfo {author} {\bibfnamefont {V.~G.}\ \bibnamefont {Pavelyev}}, \bibinfo
		{author} {\bibfnamefont {P.~H.~M.}\ \bibnamefont {van Loosdrecht}}, \bibinfo
		{author} {\bibfnamefont {M.~S.}\ \bibnamefont {Pshenichnikov}}, \bibinfo
		{author} {\bibfnamefont {D.}~\bibnamefont {Niedzialek}}, \bibinfo {author}
		{\bibfnamefont {J.}~\bibnamefont {Cornil}}, \bibinfo {author} {\bibfnamefont
			{D.}~\bibnamefont {Beljonne}}, \ and\ \bibinfo {author} {\bibfnamefont
			{R.~H.}\ \bibnamefont {Friend}},\ }\href {\doibase 10.1126/science.1217745}
	{\bibfield  {journal} {\bibinfo  {journal} {Science}\ }\textbf {\bibinfo
			{volume} {335}},\ \bibinfo {pages} {1340} (\bibinfo {year}
		{2012})}\BibitemShut {NoStop}%
	\bibitem [{\citenamefont {Hood}\ and\ \citenamefont
		{Kassal}(2016)}]{hood_entropy_2016}%
	\BibitemOpen
	\bibfield  {author} {\bibinfo {author} {\bibfnamefont {S.~N.}\ \bibnamefont
			{Hood}}\ and\ \bibinfo {author} {\bibfnamefont {I.}~\bibnamefont {Kassal}},\
	}\href {\doibase 10.1021/acs.jpclett.6b02178} {\bibfield  {journal} {\bibinfo
			{journal} {J. Phys. Chem. Lett.}\ }\textbf {\bibinfo {volume} {7}},\
		\bibinfo {pages} {4495} (\bibinfo {year} {2016})}\BibitemShut {NoStop}%
	\bibitem [{\citenamefont {Coehoorn}\ and\ \citenamefont
		{Bobbert}(2012)}]{coehoorn_effects_2012}%
	\BibitemOpen
	\bibfield  {author} {\bibinfo {author} {\bibfnamefont {R.}~\bibnamefont
			{Coehoorn}}\ and\ \bibinfo {author} {\bibfnamefont {P.~A.}\ \bibnamefont
			{Bobbert}},\ }\href {\doibase 10.1002/pssa.201228387} {\bibfield  {journal}
		{\bibinfo  {journal} {physica status solidi (a)}\ }\textbf {\bibinfo {volume}
			{209}},\ \bibinfo {pages} {2354} (\bibinfo {year} {2012})}\BibitemShut
	{NoStop}%
	\bibitem [{\citenamefont {Plenio}\ and\ \citenamefont
		{Huelga}(2008)}]{plenio_dephasing-assisted_2008}%
	\BibitemOpen
	\bibfield  {author} {\bibinfo {author} {\bibfnamefont {M.~B.}\ \bibnamefont
			{Plenio}}\ and\ \bibinfo {author} {\bibfnamefont {S.~F.}\ \bibnamefont
			{Huelga}},\ }\href {\doibase 10.1088/1367-2630/10/11/113019} {\bibfield
		{journal} {\bibinfo  {journal} {New J. Phys.}\ }\textbf {\bibinfo {volume}
			{10}},\ \bibinfo {pages} {113019} (\bibinfo {year} {2008})}\BibitemShut
	{NoStop}%
	\bibitem [{\citenamefont {Ringsmuth}\ \emph {et~al.}(2012)\citenamefont
		{Ringsmuth}, \citenamefont {Milburn},\ and\ \citenamefont
		{Stace}}]{ringsmuth_multiscale_2012}%
	\BibitemOpen
	\bibfield  {author} {\bibinfo {author} {\bibfnamefont {A.~K.}\ \bibnamefont
			{Ringsmuth}}, \bibinfo {author} {\bibfnamefont {G.~J.}\ \bibnamefont
			{Milburn}}, \ and\ \bibinfo {author} {\bibfnamefont {T.~M.}\ \bibnamefont
			{Stace}},\ }\href {\doibase 10.1038/nphys2332} {\bibfield  {journal}
		{\bibinfo  {journal} {Nature Phys}\ }\textbf {\bibinfo {volume} {8}},\
		\bibinfo {pages} {562} (\bibinfo {year} {2012})}\BibitemShut {NoStop}%
	\bibitem [{\citenamefont {Caruso}\ \emph {et~al.}(2009)\citenamefont {Caruso},
		\citenamefont {Chin}, \citenamefont {Datta}, \citenamefont {Huelga},\ and\
		\citenamefont {Plenio}}]{caruso_highly_2009}%
	\BibitemOpen
	\bibfield  {author} {\bibinfo {author} {\bibfnamefont {F.}~\bibnamefont
			{Caruso}}, \bibinfo {author} {\bibfnamefont {A.~W.}\ \bibnamefont {Chin}},
		\bibinfo {author} {\bibfnamefont {A.}~\bibnamefont {Datta}}, \bibinfo
		{author} {\bibfnamefont {S.~F.}\ \bibnamefont {Huelga}}, \ and\ \bibinfo
		{author} {\bibfnamefont {M.~B.}\ \bibnamefont {Plenio}},\ }\href {\doibase
		10.1063/1.3223548} {\bibfield  {journal} {\bibinfo  {journal} {J. Chem.
				Phys.}\ }\textbf {\bibinfo {volume} {131}},\ \bibinfo {pages} {105106}
		(\bibinfo {year} {2009})},\ \bibinfo {note} {publisher: American Institute of
		Physics}\BibitemShut {NoStop}%
	\bibitem [{\citenamefont {Caruso}\ \emph {et~al.}(2012)\citenamefont {Caruso},
		\citenamefont {Montangero}, \citenamefont {Calarco}, \citenamefont {Huelga},\
		and\ \citenamefont {Plenio}}]{caruso_coherent_2012}%
	\BibitemOpen
	\bibfield  {author} {\bibinfo {author} {\bibfnamefont {F.}~\bibnamefont
			{Caruso}}, \bibinfo {author} {\bibfnamefont {S.}~\bibnamefont {Montangero}},
		\bibinfo {author} {\bibfnamefont {T.}~\bibnamefont {Calarco}}, \bibinfo
		{author} {\bibfnamefont {S.~F.}\ \bibnamefont {Huelga}}, \ and\ \bibinfo
		{author} {\bibfnamefont {M.~B.}\ \bibnamefont {Plenio}},\ }\href {\doibase
		10.1103/PhysRevA.85.042331} {\bibfield  {journal} {\bibinfo  {journal} {Phys.
				Rev. A}\ }\textbf {\bibinfo {volume} {85}},\ \bibinfo {pages} {042331}
		(\bibinfo {year} {2012})}\BibitemShut {NoStop}%
	\bibitem [{\citenamefont {Bittner}\ and\ \citenamefont
		{Silva}(2014)}]{bittner_noise-induced_2014}%
	\BibitemOpen
	\bibfield  {author} {\bibinfo {author} {\bibfnamefont {E.~R.}\ \bibnamefont
			{Bittner}}\ and\ \bibinfo {author} {\bibfnamefont {C.}~\bibnamefont
			{Silva}},\ }\href {\doibase 10.1038/ncomms4119} {\bibfield  {journal}
		{\bibinfo  {journal} {Nat Commun}\ }\textbf {\bibinfo {volume} {5}},\
		\bibinfo {pages} {3119} (\bibinfo {year} {2014})},\ \bibinfo {note}
	{bandiera\_abtest: a Cg\_type: Nature Research Journals Number: 1
		Primary\_atype: Research Publisher: Nature Publishing Group Subject\_term:
		Quantum mechanics;Semiconductors Subject\_term\_id:
		quantum-mechanics;semiconductors}\BibitemShut {NoStop}%
	\bibitem [{\citenamefont {Gammaitoni}\ \emph {et~al.}(1998)\citenamefont
		{Gammaitoni}, \citenamefont {Hänggi}, \citenamefont {Jung},\ and\
		\citenamefont {Marchesoni}}]{gammaitoni_stochastic_1998}%
	\BibitemOpen
	\bibfield  {author} {\bibinfo {author} {\bibfnamefont {L.}~\bibnamefont
			{Gammaitoni}}, \bibinfo {author} {\bibfnamefont {P.}~\bibnamefont {Hänggi}},
		\bibinfo {author} {\bibfnamefont {P.}~\bibnamefont {Jung}}, \ and\ \bibinfo
		{author} {\bibfnamefont {F.}~\bibnamefont {Marchesoni}},\ }\href {\doibase
		10.1103/RevModPhys.70.223} {\bibfield  {journal} {\bibinfo  {journal} {Rev.
				Mod. Phys.}\ }\textbf {\bibinfo {volume} {70}},\ \bibinfo {pages} {223}
		(\bibinfo {year} {1998})},\ \bibinfo {note} {publisher: American Physical
		Society}\BibitemShut {NoStop}%
	\bibitem [{\citenamefont {Broome}\ \emph {et~al.}(2010)\citenamefont {Broome},
		\citenamefont {Fedrizzi}, \citenamefont {Lanyon}, \citenamefont {Kassal},
		\citenamefont {Aspuru-Guzik},\ and\ \citenamefont
		{White}}]{broome_discrete_2010}%
	\BibitemOpen
	\bibfield  {author} {\bibinfo {author} {\bibfnamefont {M.~A.}\ \bibnamefont
			{Broome}}, \bibinfo {author} {\bibfnamefont {A.}~\bibnamefont {Fedrizzi}},
		\bibinfo {author} {\bibfnamefont {B.~P.}\ \bibnamefont {Lanyon}}, \bibinfo
		{author} {\bibfnamefont {I.}~\bibnamefont {Kassal}}, \bibinfo {author}
		{\bibfnamefont {A.}~\bibnamefont {Aspuru-Guzik}}, \ and\ \bibinfo {author}
		{\bibfnamefont {A.~G.}\ \bibnamefont {White}},\ }\href {\doibase
		10.1103/PhysRevLett.104.153602} {\bibfield  {journal} {\bibinfo  {journal}
			{Phys. Rev. Lett.}\ }\textbf {\bibinfo {volume} {104}},\ \bibinfo {pages}
		{153602} (\bibinfo {year} {2010})},\ \bibinfo {note} {publisher: American
		Physical Society}\BibitemShut {NoStop}%
	\bibitem [{\citenamefont {Biggerstaff}\ \emph {et~al.}(2016)\citenamefont
		{Biggerstaff}, \citenamefont {Heilmann}, \citenamefont {Zecevik},
		\citenamefont {Gräfe}, \citenamefont {Broome}, \citenamefont {Fedrizzi},
		\citenamefont {Nolte}, \citenamefont {Szameit}, \citenamefont {White},\ and\
		\citenamefont {Kassal}}]{biggerstaff_enhancing_2016}%
	\BibitemOpen
	\bibfield  {author} {\bibinfo {author} {\bibfnamefont {D.~N.}\ \bibnamefont
			{Biggerstaff}}, \bibinfo {author} {\bibfnamefont {R.}~\bibnamefont
			{Heilmann}}, \bibinfo {author} {\bibfnamefont {A.~A.}\ \bibnamefont
			{Zecevik}}, \bibinfo {author} {\bibfnamefont {M.}~\bibnamefont {Gräfe}},
		\bibinfo {author} {\bibfnamefont {M.~A.}\ \bibnamefont {Broome}}, \bibinfo
		{author} {\bibfnamefont {A.}~\bibnamefont {Fedrizzi}}, \bibinfo {author}
		{\bibfnamefont {S.}~\bibnamefont {Nolte}}, \bibinfo {author} {\bibfnamefont
			{A.}~\bibnamefont {Szameit}}, \bibinfo {author} {\bibfnamefont {A.~G.}\
			\bibnamefont {White}}, \ and\ \bibinfo {author} {\bibfnamefont
			{I.}~\bibnamefont {Kassal}},\ }\href {\doibase 10.1038/ncomms11282}
	{\bibfield  {journal} {\bibinfo  {journal} {Nature Communications}\ }\textbf
		{\bibinfo {volume} {7}},\ \bibinfo {pages} {11282} (\bibinfo {year}
		{2016})}\BibitemShut {NoStop}%
	\bibitem [{\citenamefont {Wang}\ \emph {et~al.}(2018)\citenamefont {Wang},
		\citenamefont {Tao}, \citenamefont {Ai}, \citenamefont {Xin}, \citenamefont
		{Lambert}, \citenamefont {Ruan}, \citenamefont {Cheng}, \citenamefont {Nori},
		\citenamefont {Deng},\ and\ \citenamefont {Long}}]{wang_efficient_2018}%
	\BibitemOpen
	\bibfield  {author} {\bibinfo {author} {\bibfnamefont {B.-X.}\ \bibnamefont
			{Wang}}, \bibinfo {author} {\bibfnamefont {M.-J.}\ \bibnamefont {Tao}},
		\bibinfo {author} {\bibfnamefont {Q.}~\bibnamefont {Ai}}, \bibinfo {author}
		{\bibfnamefont {T.}~\bibnamefont {Xin}}, \bibinfo {author} {\bibfnamefont
			{N.}~\bibnamefont {Lambert}}, \bibinfo {author} {\bibfnamefont
			{D.}~\bibnamefont {Ruan}}, \bibinfo {author} {\bibfnamefont {Y.-C.}\
			\bibnamefont {Cheng}}, \bibinfo {author} {\bibfnamefont {F.}~\bibnamefont
			{Nori}}, \bibinfo {author} {\bibfnamefont {F.-G.}\ \bibnamefont {Deng}}, \
		and\ \bibinfo {author} {\bibfnamefont {G.-L.}\ \bibnamefont {Long}},\ }\href
	{\doibase 10.1038/s41534-018-0102-2} {\bibfield  {journal} {\bibinfo
			{journal} {npj Quantum Inf}\ }\textbf {\bibinfo {volume} {4}},\ \bibinfo
		{pages} {1} (\bibinfo {year} {2018})},\ \bibinfo {note} {bandiera\_abtest: a
		Cc\_license\_type: cc\_by Cg\_type: Nature Research Journals Number: 1
		Primary\_atype: Research Publisher: Nature Publishing Group Subject\_term:
		Quantum information;Quantum optics Subject\_term\_id:
		quantum-information;quantum-optics}\BibitemShut {NoStop}%
	\bibitem [{\citenamefont {Jiao}\ \emph {et~al.}(2021)\citenamefont {Jiao},
		\citenamefont {Jiao}, \citenamefont {Gao}, \citenamefont {Gao}, \citenamefont
		{Zhou}, \citenamefont {Zhou}, \citenamefont {Wang}, \citenamefont {Wang},
		\citenamefont {Ren}, \citenamefont {Ren}, \citenamefont {Xu}, \citenamefont
		{Xu}, \citenamefont {Qiao}, \citenamefont {Qiao}, \citenamefont {Wang},
		\citenamefont {Wang}, \citenamefont {Jin},\ and\ \citenamefont
		{Jin}}]{jiao_two-dimensional_2021}%
	\BibitemOpen
	\bibfield  {author} {\bibinfo {author} {\bibfnamefont {Z.-Q.}\ \bibnamefont
			{Jiao}}, \bibinfo {author} {\bibfnamefont {Z.-Q.}\ \bibnamefont {Jiao}},
		\bibinfo {author} {\bibfnamefont {J.}~\bibnamefont {Gao}}, \bibinfo {author}
		{\bibfnamefont {J.}~\bibnamefont {Gao}}, \bibinfo {author} {\bibfnamefont
			{W.-H.}\ \bibnamefont {Zhou}}, \bibinfo {author} {\bibfnamefont {W.-H.}\
			\bibnamefont {Zhou}}, \bibinfo {author} {\bibfnamefont {X.-W.}\ \bibnamefont
			{Wang}}, \bibinfo {author} {\bibfnamefont {X.-W.}\ \bibnamefont {Wang}},
		\bibinfo {author} {\bibfnamefont {R.-J.}\ \bibnamefont {Ren}}, \bibinfo
		{author} {\bibfnamefont {R.-J.}\ \bibnamefont {Ren}}, \bibinfo {author}
		{\bibfnamefont {X.-Y.}\ \bibnamefont {Xu}}, \bibinfo {author} {\bibfnamefont
			{X.-Y.}\ \bibnamefont {Xu}}, \bibinfo {author} {\bibfnamefont {L.-F.}\
			\bibnamefont {Qiao}}, \bibinfo {author} {\bibfnamefont {L.-F.}\ \bibnamefont
			{Qiao}}, \bibinfo {author} {\bibfnamefont {Y.}~\bibnamefont {Wang}}, \bibinfo
		{author} {\bibfnamefont {Y.}~\bibnamefont {Wang}}, \bibinfo {author}
		{\bibfnamefont {X.-M.}\ \bibnamefont {Jin}}, \ and\ \bibinfo {author}
		{\bibfnamefont {X.-M.}\ \bibnamefont {Jin}},\ }\href {\doibase
		10.1364/OPTICA.425879} {\bibfield  {journal} {\bibinfo  {journal} {Optica,
				OPTICA}\ }\textbf {\bibinfo {volume} {8}},\ \bibinfo {pages} {1129} (\bibinfo
		{year} {2021})},\ \bibinfo {note} {publisher: Optical Society of
		America}\BibitemShut {NoStop}%
	\bibitem [{\citenamefont {Owens}\ \emph {et~al.}(2011)\citenamefont {Owens},
		\citenamefont {Broome}, \citenamefont {Biggerstaff}, \citenamefont {Goggin},
		\citenamefont {Fedrizzi}, \citenamefont {Linjordet}, \citenamefont {Ams},
		\citenamefont {Marshall}, \citenamefont {Twamley}, \citenamefont {Withford},\
		and\ \citenamefont {White}}]{owens_two-photon_2011}%
	\BibitemOpen
	\bibfield  {author} {\bibinfo {author} {\bibfnamefont {J.~O.}\ \bibnamefont
			{Owens}}, \bibinfo {author} {\bibfnamefont {M.~A.}\ \bibnamefont {Broome}},
		\bibinfo {author} {\bibfnamefont {D.~N.}\ \bibnamefont {Biggerstaff}},
		\bibinfo {author} {\bibfnamefont {M.~E.}\ \bibnamefont {Goggin}}, \bibinfo
		{author} {\bibfnamefont {A.}~\bibnamefont {Fedrizzi}}, \bibinfo {author}
		{\bibfnamefont {T.}~\bibnamefont {Linjordet}}, \bibinfo {author}
		{\bibfnamefont {M.}~\bibnamefont {Ams}}, \bibinfo {author} {\bibfnamefont
			{G.~D.}\ \bibnamefont {Marshall}}, \bibinfo {author} {\bibfnamefont
			{J.}~\bibnamefont {Twamley}}, \bibinfo {author} {\bibfnamefont {M.~J.}\
			\bibnamefont {Withford}}, \ and\ \bibinfo {author} {\bibfnamefont {A.~G.}\
			\bibnamefont {White}},\ }\href {\doibase 10.1088/1367-2630/13/7/075003}
	{\bibfield  {journal} {\bibinfo  {journal} {New J. Phys.}\ }\textbf {\bibinfo
			{volume} {13}},\ \bibinfo {pages} {075003} (\bibinfo {year} {2011})},\
	\bibinfo {note} {publisher: IOP Publishing}\BibitemShut {NoStop}%
	\bibitem [{\citenamefont {Motes}\ \emph {et~al.}(2016)\citenamefont {Motes},
		\citenamefont {Gilchrist},\ and\ \citenamefont {Rohde}}]{motes_quantum_2016}%
	\BibitemOpen
	\bibfield  {author} {\bibinfo {author} {\bibfnamefont {K.~R.}\ \bibnamefont
			{Motes}}, \bibinfo {author} {\bibfnamefont {A.}~\bibnamefont {Gilchrist}}, \
		and\ \bibinfo {author} {\bibfnamefont {P.~P.}\ \bibnamefont {Rohde}},\ }\href
	{\doibase 10.1038/srep19864} {\bibfield  {journal} {\bibinfo  {journal} {Sci
				Rep}\ }\textbf {\bibinfo {volume} {6}},\ \bibinfo {pages} {19864} (\bibinfo
		{year} {2016})},\ \bibinfo {note} {bandiera\_abtest: a Cc\_license\_type:
		cc\_by Cg\_type: Nature Research Journals Number: 1 Primary\_atype: Research
		Publisher: Nature Publishing Group Subject\_term: Quantum
		simulation;Theoretical physics Subject\_term\_id:
		quantum-simulation;theoretical-physics}\BibitemShut {NoStop}%
	\bibitem [{\citenamefont {Dalla~Pozza}\ and\ \citenamefont
		{Caruso}(2020)}]{dalla_pozza_quantum_2020}%
	\BibitemOpen
	\bibfield  {author} {\bibinfo {author} {\bibfnamefont {N.}~\bibnamefont
			{Dalla~Pozza}}\ and\ \bibinfo {author} {\bibfnamefont {F.}~\bibnamefont
			{Caruso}},\ }\href {\doibase 10.1016/j.physleta.2019.126195} {\bibfield
		{journal} {\bibinfo  {journal} {Physics Letters A}\ }\textbf {\bibinfo
			{volume} {384}},\ \bibinfo {pages} {126195} (\bibinfo {year}
		{2020})}\BibitemShut {NoStop}%
	\bibitem [{\citenamefont {Rebentrost}\ and\ \citenamefont
		{Lloyd}(2018)}]{rebentrost_quantum_2018}%
	\BibitemOpen
	\bibfield  {author} {\bibinfo {author} {\bibfnamefont {P.}~\bibnamefont
			{Rebentrost}}\ and\ \bibinfo {author} {\bibfnamefont {S.}~\bibnamefont
			{Lloyd}},\ }\href {http://arxiv.org/abs/1811.03975} {\bibfield  {journal}
		{\bibinfo  {journal} {arXiv:1811.03975 [quant-ph]}\ } (\bibinfo {year}
		{2018})},\ \bibinfo {note} {arXiv: 1811.03975}\BibitemShut {NoStop}%
	\bibitem [{\citenamefont {Siloi}\ \emph {et~al.}(2017)\citenamefont {Siloi},
		\citenamefont {Benedetti}, \citenamefont {Piccinini}, \citenamefont {Piilo},
		\citenamefont {Maniscalco}, \citenamefont {Paris},\ and\ \citenamefont
		{Bordone}}]{siloi_noisy_2017}%
	\BibitemOpen
	\bibfield  {author} {\bibinfo {author} {\bibfnamefont {I.}~\bibnamefont
			{Siloi}}, \bibinfo {author} {\bibfnamefont {C.}~\bibnamefont {Benedetti}},
		\bibinfo {author} {\bibfnamefont {E.}~\bibnamefont {Piccinini}}, \bibinfo
		{author} {\bibfnamefont {J.}~\bibnamefont {Piilo}}, \bibinfo {author}
		{\bibfnamefont {S.}~\bibnamefont {Maniscalco}}, \bibinfo {author}
		{\bibfnamefont {M.~G.~A.}\ \bibnamefont {Paris}}, \ and\ \bibinfo {author}
		{\bibfnamefont {P.}~\bibnamefont {Bordone}},\ }\href {\doibase
		10.1103/PhysRevA.95.022106} {\bibfield  {journal} {\bibinfo  {journal} {Phys.
				Rev. A}\ }\textbf {\bibinfo {volume} {95}},\ \bibinfo {pages} {022106}
		(\bibinfo {year} {2017})},\ \bibinfo {note} {publisher: American Physical
		Society}\BibitemShut {NoStop}%
	\bibitem [{\citenamefont {Uchiyama}\ \emph {et~al.}(2018)\citenamefont
		{Uchiyama}, \citenamefont {Munro},\ and\ \citenamefont
		{Nemoto}}]{uchiyama_environmental_2018}%
	\BibitemOpen
	\bibfield  {author} {\bibinfo {author} {\bibfnamefont {C.}~\bibnamefont
			{Uchiyama}}, \bibinfo {author} {\bibfnamefont {W.~J.}\ \bibnamefont {Munro}},
		\ and\ \bibinfo {author} {\bibfnamefont {K.}~\bibnamefont {Nemoto}},\ }\href
	{\doibase 10.1038/s41534-018-0079-x} {\bibfield  {journal} {\bibinfo
			{journal} {npj Quantum Inf}\ }\textbf {\bibinfo {volume} {4}},\ \bibinfo
		{pages} {1} (\bibinfo {year} {2018})}\BibitemShut {NoStop}%
	\bibitem [{\citenamefont {Wiseman}(1996)}]{wiseman_quantum_1996}%
	\BibitemOpen
	\bibfield  {author} {\bibinfo {author} {\bibfnamefont {H.~M.}\ \bibnamefont
			{Wiseman}},\ }\href {\doibase 10.1088/1355-5111/8/1/015} {\bibfield
		{journal} {\bibinfo  {journal} {Quantum Semiclass. Opt.}\ }\textbf {\bibinfo
			{volume} {8}},\ \bibinfo {pages} {205} (\bibinfo {year} {1996})}\BibitemShut
	{NoStop}%
	\bibitem [{\citenamefont {Daley}(2014)}]{daley_quantum_2014}%
	\BibitemOpen
	\bibfield  {author} {\bibinfo {author} {\bibfnamefont {A.~J.}\ \bibnamefont
			{Daley}},\ }\href {\doibase 10.1080/00018732.2014.933502} {\bibfield
		{journal} {\bibinfo  {journal} {Advances in Physics}\ }\textbf {\bibinfo
			{volume} {63}},\ \bibinfo {pages} {77} (\bibinfo {year} {2014})},\ \bibinfo
	{note} {arXiv: 1405.6694}\BibitemShut {NoStop}%
	\bibitem [{\citenamefont {Gardiner}\ and\ \citenamefont
		{Collett}(1985)}]{gardiner_input_1985}%
	\BibitemOpen
	\bibfield  {author} {\bibinfo {author} {\bibfnamefont {C.~W.}\ \bibnamefont
			{Gardiner}}\ and\ \bibinfo {author} {\bibfnamefont {M.~J.}\ \bibnamefont
			{Collett}},\ }\href {\doibase 10.1103/PhysRevA.31.3761} {\bibfield  {journal}
		{\bibinfo  {journal} {Phys. Rev. A}\ }\textbf {\bibinfo {volume} {31}},\
		\bibinfo {pages} {3761} (\bibinfo {year} {1985})}\BibitemShut {NoStop}%
	\bibitem [{\citenamefont {Zeb}\ \emph {et~al.}(2020)\citenamefont {Zeb},
		\citenamefont {Kirton},\ and\ \citenamefont {Keeling}}]{zeb_incoherent_2020}%
	\BibitemOpen
	\bibfield  {author} {\bibinfo {author} {\bibfnamefont {M.~A.}\ \bibnamefont
			{Zeb}}, \bibinfo {author} {\bibfnamefont {P.~G.}\ \bibnamefont {Kirton}}, \
		and\ \bibinfo {author} {\bibfnamefont {J.}~\bibnamefont {Keeling}},\ }\href
	{http://arxiv.org/abs/2004.09790} {\bibfield  {journal} {\bibinfo  {journal}
			{arXiv:2004.09790 [cond-mat]}\ } (\bibinfo {year} {2020})},\ \bibinfo {note}
	{arXiv: 2004.09790}\BibitemShut {NoStop}%
	\bibitem [{\citenamefont {Strashko}\ \emph {et~al.}(2018)\citenamefont
		{Strashko}, \citenamefont {Kirton},\ and\ \citenamefont
		{Keeling}}]{strashko_organic_2018}%
	\BibitemOpen
	\bibfield  {author} {\bibinfo {author} {\bibfnamefont {A.}~\bibnamefont
			{Strashko}}, \bibinfo {author} {\bibfnamefont {P.}~\bibnamefont {Kirton}}, \
		and\ \bibinfo {author} {\bibfnamefont {J.}~\bibnamefont {Keeling}},\ }\href
	{\doibase 10.1103/PhysRevLett.121.193601} {\bibfield  {journal} {\bibinfo
			{journal} {Phys. Rev. Lett.}\ }\textbf {\bibinfo {volume} {121}},\ \bibinfo
		{pages} {193601} (\bibinfo {year} {2018})}\BibitemShut {NoStop}%
	\bibitem [{\citenamefont {Sakurai}(1994)}]{Sakurai:1167961}%
	\BibitemOpen
	\bibfield  {author} {\bibinfo {author} {\bibfnamefont {J.~J.}\ \bibnamefont
			{Sakurai}},\ }\href {https://cds.cern.ch/record/1167961} {\emph {\bibinfo
			{title} {{Modern quantum mechanics; rev. ed.}}}}\ (\bibinfo  {publisher}
	{Addison-Wesley},\ \bibinfo {address} {Reading, MA},\ \bibinfo {year}
	{1994})\BibitemShut {NoStop}%
	\bibitem [{\citenamefont {D'Errico}\ \emph {et~al.}(2013)\citenamefont
		{D'Errico}, \citenamefont {Moratti}, \citenamefont {Lucioni}, \citenamefont
		{Tanzi}, \citenamefont {Deissler}, \citenamefont {Inguscio}, \citenamefont
		{Modugno}, \citenamefont {Plenio},\ and\ \citenamefont
		{Caruso}}]{derrico_quantum_2013}%
	\BibitemOpen
	\bibfield  {author} {\bibinfo {author} {\bibfnamefont {C.}~\bibnamefont
			{D'Errico}}, \bibinfo {author} {\bibfnamefont {M.}~\bibnamefont {Moratti}},
		\bibinfo {author} {\bibfnamefont {E.}~\bibnamefont {Lucioni}}, \bibinfo
		{author} {\bibfnamefont {L.}~\bibnamefont {Tanzi}}, \bibinfo {author}
		{\bibfnamefont {B.}~\bibnamefont {Deissler}}, \bibinfo {author}
		{\bibfnamefont {M.}~\bibnamefont {Inguscio}}, \bibinfo {author}
		{\bibfnamefont {G.}~\bibnamefont {Modugno}}, \bibinfo {author} {\bibfnamefont
			{M.~B.}\ \bibnamefont {Plenio}}, \ and\ \bibinfo {author} {\bibfnamefont
			{F.}~\bibnamefont {Caruso}},\ }\href {\doibase 10.1088/1367-2630/15/4/045007}
	{\bibfield  {journal} {\bibinfo  {journal} {New J. Phys.}\ }\textbf {\bibinfo
			{volume} {15}},\ \bibinfo {pages} {045007} (\bibinfo {year} {2013})},\
	\bibinfo {note} {publisher: IOP Publishing}\BibitemShut {NoStop}%
	\bibitem [{\citenamefont {Caruso}\ \emph {et~al.}(2016)\citenamefont {Caruso},
		\citenamefont {Crespi}, \citenamefont {Ciriolo}, \citenamefont {Sciarrino},\
		and\ \citenamefont {Osellame}}]{caruso_fast_2016}%
	\BibitemOpen
	\bibfield  {author} {\bibinfo {author} {\bibfnamefont {F.}~\bibnamefont
			{Caruso}}, \bibinfo {author} {\bibfnamefont {A.}~\bibnamefont {Crespi}},
		\bibinfo {author} {\bibfnamefont {A.~G.}\ \bibnamefont {Ciriolo}}, \bibinfo
		{author} {\bibfnamefont {F.}~\bibnamefont {Sciarrino}}, \ and\ \bibinfo
		{author} {\bibfnamefont {R.}~\bibnamefont {Osellame}},\ }\href {\doibase
		10.1038/ncomms11682} {\bibfield  {journal} {\bibinfo  {journal} {Nat Commun}\
		}\textbf {\bibinfo {volume} {7}},\ \bibinfo {pages} {11682} (\bibinfo {year}
		{2016})},\ \bibinfo {note} {bandiera\_abtest: a Cc\_license\_type: cc\_by
		Cg\_type: Nature Research Journals Number: 1 Primary\_atype: Research
		Publisher: Nature Publishing Group Subject\_term: Biophotonics;Quantum
		mechanics;Theoretical physics Subject\_term\_id:
		biophotonics;quantum-mechanics;theoretical-physics}\BibitemShut {NoStop}%
	\bibitem [{\citenamefont {Anderson}(1958)}]{anderson_absence_1958}%
	\BibitemOpen
	\bibfield  {author} {\bibinfo {author} {\bibfnamefont {P.~W.}\ \bibnamefont
			{Anderson}},\ }\href {\doibase 10.1103/PhysRev.109.1492} {\bibfield
		{journal} {\bibinfo  {journal} {Phys. Rev.}\ }\textbf {\bibinfo {volume}
			{109}},\ \bibinfo {pages} {1492} (\bibinfo {year} {1958})},\ \bibinfo {note}
	{publisher: American Physical Society}\BibitemShut {NoStop}%
	\bibitem [{\citenamefont {Wiseman}\ and\ \citenamefont
		{Doherty}(2005)}]{wiseman_optimal_2005}%
	\BibitemOpen
	\bibfield  {author} {\bibinfo {author} {\bibfnamefont {H.~M.}\ \bibnamefont
			{Wiseman}}\ and\ \bibinfo {author} {\bibfnamefont {A.~C.}\ \bibnamefont
			{Doherty}},\ }\href {\doibase 10.1103/PhysRevLett.94.070405} {\bibfield
		{journal} {\bibinfo  {journal} {Phys. Rev. Lett.}\ }\textbf {\bibinfo
			{volume} {94}},\ \bibinfo {pages} {070405} (\bibinfo {year}
		{2005})}\BibitemShut {NoStop}%
	\bibitem [{\citenamefont {Wiseman}\ and\ \citenamefont
		{Milburn}(1993)}]{wiseman_interpretation_1993}%
	\BibitemOpen
	\bibfield  {author} {\bibinfo {author} {\bibfnamefont {H.~M.}\ \bibnamefont
			{Wiseman}}\ and\ \bibinfo {author} {\bibfnamefont {G.~J.}\ \bibnamefont
			{Milburn}},\ }\href {\doibase 10.1103/PhysRevA.47.1652} {\bibfield  {journal}
		{\bibinfo  {journal} {Phys. Rev. A}\ }\textbf {\bibinfo {volume} {47}},\
		\bibinfo {pages} {1652} (\bibinfo {year} {1993})},\ \bibinfo {note}
	{publisher: American Physical Society}\BibitemShut {NoStop}%
	\bibitem [{\citenamefont {Keys}\ and\ \citenamefont
		{Wehr}(2020)}]{keys_poisson_2020}%
	\BibitemOpen
	\bibfield  {author} {\bibinfo {author} {\bibfnamefont {D.}~\bibnamefont
			{Keys}}\ and\ \bibinfo {author} {\bibfnamefont {J.}~\bibnamefont {Wehr}},\
	}\href {\doibase 10.1063/1.5133974} {\bibfield  {journal} {\bibinfo
			{journal} {J. Math. Phys.}\ }\textbf {\bibinfo {volume} {61}},\ \bibinfo
		{pages} {032101} (\bibinfo {year} {2020})},\ \bibinfo {note} {publisher:
		American Institute of Physics}\BibitemShut {NoStop}%
	\bibitem [{\citenamefont {Schwartz}\ \emph {et~al.}(2007)\citenamefont
		{Schwartz}, \citenamefont {Bartal}, \citenamefont {Fishman},\ and\
		\citenamefont {Segev}}]{schwartz_transport_2007}%
	\BibitemOpen
	\bibfield  {author} {\bibinfo {author} {\bibfnamefont {T.}~\bibnamefont
			{Schwartz}}, \bibinfo {author} {\bibfnamefont {G.}~\bibnamefont {Bartal}},
		\bibinfo {author} {\bibfnamefont {S.}~\bibnamefont {Fishman}}, \ and\
		\bibinfo {author} {\bibfnamefont {M.}~\bibnamefont {Segev}},\ }\href
	{\doibase 10.1038/nature05623} {\bibfield  {journal} {\bibinfo  {journal}
			{Nature}\ }\textbf {\bibinfo {volume} {446}},\ \bibinfo {pages} {52}
		(\bibinfo {year} {2007})},\ \bibinfo {note} {bandiera\_abtest: a Cg\_type:
		Nature Research Journals Number: 7131 Primary\_atype: Research Publisher:
		Nature Publishing Group}\BibitemShut {NoStop}%
	\bibitem [{\citenamefont {Taylor}\ and\ \citenamefont
		{Kassal}(2018)}]{taylor_generalised_2018}%
	\BibitemOpen
	\bibfield  {author} {\bibinfo {author} {\bibfnamefont {N.~B.}\ \bibnamefont
			{Taylor}}\ and\ \bibinfo {author} {\bibfnamefont {I.}~\bibnamefont
			{Kassal}},\ }\href {\doibase 10.1039/C8SC00053K} {\bibfield  {journal}
		{\bibinfo  {journal} {Chem. Sci.}\ }\textbf {\bibinfo {volume} {9}},\
		\bibinfo {pages} {2942} (\bibinfo {year} {2018})}\BibitemShut {NoStop}%
\end{thebibliography}

%

\newpage
\begin{widetext}
\appendix

\section{On-site dephasing master equation}

Here we consider the case of white noise dephasing, where $\epsilon(t)$ is a white noise process. Formally, this white noise arises form coupling to many bath degrees of freedom, and making the Born-Markov approximation for the bath. In this case the magnitude of the white noise process can then be related to the thermal properties of the bath e.g. in ref \cite{gardiner_input_1985,taylor_generalised_2018}. For simplicity here we simply consider a classical while noise process. 

A classical white noise process can be informally understood as the derivative of a Weiner process: $\epsilon(t) = \sqrt{\gamma} dW(t)/dt$. This can be see as follows: Choose $\epsilon_j(t) dt = \sqrt{\gamma_j}dW_j(t)$, giving $\mathcal{E}(\epsilon_j(t)  \epsilon_j(t') )dt^2 = \mathcal{E}(dW(t) dW(t'))\gamma = \delta_{tt'} dt \gamma$, (where $dW(t)$ is an Ito increment) Therefore $\mathcal{E}(\epsilon(t)\epsilon(t')) = \frac{\gamma \delta_{tt'}}{dt} = \gamma \delta(t-t')$. This last expression, $\delta_{tt'}/{dt} = \delta(t-t')$ can be simply understood by considering discrete time intervals $\delta t$, with the kronoker-delta evaluated between interval beginning at $t$ and $t'$ (both beginning at the beginning of some discrete period) in the Ito frame.

Consider the infinitesimal evolution of an initial state $\rho \equiv \rho(t)$. Writing the white noise increment as a Weiner process, and keeping terms to first order in $dt$ (second order in $dW$) we have, 

\begin{eqnarray}
\rho(t + dt) &=& e^{-iHdt}\rho e^{iHdt} \nl
 &=& e^{-iH_0 dt -i \sum_j\sqrt{\gamma_j} dW_j\sigmaz_j}\rho e^{iH_0 dt + i\sum_j\sqrt{\gamma_j} dW_j\sigmaz_j}\nl
&=&\left [1 - i H_0dt  - \sum_j\left(i\sqrt{\gamma_j} dW_j \sigmaz_j  +\frac12\gamma_j dt\right) - \frac12 \sum_{j\neq k}\sqrt{\gamma_j\gamma_k}dW_jdW_k\sigmaz_j\sigmaz_k \right] \rho \nl
& & \times \left[1 + i H_0dt + \sum_j\left(i\sqrt{\gamma_j} dW_j \sigmaz_j  -\frac12\gamma_j dt\right) -\frac12 \sum_{j\neq k} \sqrt{\gamma_j\gamma_k}dW_jdW_k \sigmaz_j \sigmaz_k\right] \nl
&=& \rho - i[H_0, \rho] dt + \sum_j (-\gamma_j dt \rho  + \gamma_j\sigmaz_j \rho \sigmaz_j - i\sqrt{\gamma_j}[\sigmaz_j, \rho] dW_j) + \sum_{k\neq j}\sqrt{\gamma_j \gamma_k}\left[\sigmaz_j \rho \sigmaz_k  - \frac12\{\sigmaz_j \sigmaz_k, \rho\}\right] dW_j dW_k  \nl
\end{eqnarray}
where the $\{\cdot, \cdot\}$ in the last line is the anti-commutator bracket. 
Different realisations of $dW_j$ correspond to different quantum trajectories taken by the state~\cite{wiseman_quantum_1996}. The stochastic master equation above preserves the purity, $\Trace{\rho(t + dt)} = \Trace{\rho}$, and therefore pure states remain pure. Taking the average of the noise process using $\mathcal{E}(dW_j)= \mathcal{E}(dW_jdW_k)= 0$ (for $j\neq k$), and noting that $\rho = \frac12({\sigmaz_j}^2 \rho + \rho{\sigmaz_j}^2 )$, we arrive at the dephasing induced master equation:
\begin{eqnarray}
\dot{\rho} = -i[H_0, \rho] + \sum_j \gamma_j \mathcal{D}[\sigmaz_j] \rho 
\end{eqnarray}

A similar calculation, with $\mathcal{E}[\delta_{jk}(t)\delta_{j'k'}(t')] = \delta_{jj'}\delta_{kk'}\Gamma_{jk}\delta(t -t')$ is used to derive Eq.~\eqref{eq:strong_deph}.

\section{Dephasing as jump localisation}

Here we show the dephasing master equation is entirely equivalent to continuous weak quantum measurement of the localisation. In a time step $dt$, we will consider the quantum state is projected in the $\sigmaz$ basis with probability $\gamma dt$. With complimentary probability $1 - \gamma dt$, noting happens.

The probabilities of which state is projected are given by the Born rule, $P_{+1} = \mbox{Tr}[\rho \ket{1}\bra{1}]$, and $P_{-1} = \mbox{Tr}[\rho \ket{-1}\bra{-1}]$, with $P_i$ the probability of projecting into state $i = \pm 1$. If an initial state is projected into a $\sigmaz$ eigenbasis, the normalised state is $\ketbra{i}{i}\rho\ketbra{i}{i}  / P_i = \ketbra{i}{i}$.

Under this random measurement process, the resulting evolution over a $dt$ time step is, 
\begin{eqnarray}
\rho(t + dt) &=& (1- \gamma dt)\rho + \gamma dt \left(P_{+1}\ketbra{+1}{+1} + P_{-1} \ketbra{-1}{-1} \right) \nl
&=& (1- \gamma dt)\rho + \gamma dt \left( \ketbra{+1}{1+}\rho \ketbra{+1}{+1} + \ketbra{-1}{-1}\rho \ketbra{-1}{-1} \right)
\end{eqnarray}
where $\ketbra{\pm1}{\pm1} = \frac12(\mathds{I}_2 \pm\sigmaz )$. Expanding this in the above expression gives,
\begin{eqnarray}
d\rho &=& \rho(t + dt ) - \rho =  - \gamma dt \rho + \gamma dt\frac14 \left[2 \sigmaz\rho\sigmaz   + 2 \rho \right] \nl
&=& \frac{\gamma dt}{2}\left[\sigmaz \rho \sigma\ - \frac12 \left( \sigmaz\rho + \rho\sigmaz \right)\right]
\end{eqnarray}
which is exactly the on site dephased master equation up to a factor of $1/2$. Hence the dephased master equation admits a projective Poisson measurement process, with unknown measurement results. One can then study the quantum trajectories in the measurement process to understand the jumps in the dephasing induced transport. The Poisson process for incoherent hops can be derived in a similar way. E.g, with probability $\Gamma dt$ a jump is made, and incoherently sum the result normalised states 

\end{widetext}

\end{document}